\documentclass[11pt]{article}
\pdfoutput=1

\usepackage{graphics, color,soul}
\usepackage{graphicx}
\usepackage{amssymb}

\usepackage{lmodern,mathtools}

\usepackage{booktabs}
\usepackage[english]{babel}
\usepackage{amsmath,amssymb,amsbsy,amstext, amsthm, simplewick}
\usepackage{hyperref}
\usepackage{tikz}
\usepackage{cite}

\usetikzlibrary{decorations.pathmorphing,shapes.misc}
\tikzset{snake it/.style={decorate, decoration=snake}}
\tikzset{cross/.style={cross out, draw=black, minimum size=2*(#1-\pgflinewidth), inner sep=0pt, outer sep=0pt},
cross/.default={1pt}}

\usepackage{amsfonts}
\usepackage{amssymb}
\usepackage{upgreek}
\usepackage{simplewick}
 \usepackage{exscale,relsize}
\usepackage{mathtools}
\usepackage{comment}

\usepackage[margin=1cm,labelfont={sf,bf,scriptsize},textfont={sf,scriptsize}]{caption}

\usepackage{colortbl}
\definecolor{lightgreen}{cmyk}{0.2, 0, 0.2, 0.2}
\definecolor{lightgray}{cmyk}{0.1,0.2,0,0.1}
\definecolor{lightgray2}{cmyk}{0.1,0.1,0,0.1}

\makeatletter
\newlength{\apb@width}
\newcommand{\autoparbox}[2][c]{\settowidth{\apb@width}{#2}\parbox[#1]{\apb@width}{#2}}

\makeatother

\numberwithin{equation}{section}

\def\beq{\begin{equation}}
\def\eeq{\end{equation}}

\def\bea{\begin{eqnarray}}
\def\eea{\end{eqnarray}}

\def\beq{\begin{equation}}
\def\eeq{\end{equation}}
\def\be{\begin{equation}}
\def\ee{\end{equation}}
\def\bea{\begin{eqnarray}}
\def\eea{\end{eqnarray}}

\def\epsn{\epsilon}

\def\0{{\vec{0}}}

\DeclareRobustCommand{\SkipTocEntry}[4]{}

\def\t{\tau}

\def\beq{\begin{equation}}
\def\eeq{\end{equation}}

\def\ba#1\ea{\begin{align}#1\end{align}}
\def\bg#1\eg{\begin{gather}#1\end{gather}}
\newcommand{\bseq}{\begin{subequations}}
\newcommand{\eseq}{\end{subequations}}

\renewcommand{\t}{\Tilde}

\newcommand{\tr}{{\text {tr}}}
\newcommand{\mc}{\mathcal}

\DeclareSymbolFont{extraup}{U}{zavm}{m}{n}
\DeclareMathSymbol{\varheart}{\mathalpha}{extraup}{86}
\DeclareMathSymbol{\vardiamond}{\mathalpha}{extraup}{87}

\def\({\left(}
\def\){\right)}
\def\[{\left[}
\def\]{\right]}

\begin{document}

\begin{titlepage}

\setcounter{page}{1} \baselineskip=15.5pt \thispagestyle{empty}

\vbox{\baselineskip14pt
}
{~~~~~~~~~~~~~~~~~~~~~~~~~~~~~~~~~~~~
~~~~~~~~~~~~~~~~~~~~~~~~~~~~~~~~~~
~~~~~~~~~~~ }

\bigskip\
\hbox{CALT-TH-2019--031}
\vspace{1cm}
\begin{center}
{\fontsize{19}{36}\selectfont  
{
$T\bar T$ and EE, with implications for (A)dS subregion encodings 
}
}
\end{center}

\vspace{0.6cm}

\begin{center}
Aitor Lewkowycz$^{1}$, Junyu Liu$^{2,3}$, Eva Silverstein$^1$, Gonzalo Torroba$^4$
\end{center}


\begin{center}
\vskip 8pt

\textsl{
\emph{$^1$Stanford Institute for Theoretical Physics, Stanford University, Stanford, CA 94306, USA}}
\vskip 7pt
\textsl{\emph{$^2$Walter Burke Institute for Theoretical Physics, California Institute of Technology, Pasadena, CA 91125, USA}}
\vskip 7pt
\textsl{\emph{$^3$Institute for Quantum Information and Matter, California Institute of Technology, Pasadena, CA 91125, USA}}
\vskip 7pt
\textsl{ \emph{$^4$Centro At\'omico Bariloche and CONICET, Bariloche, Argentina}}

\end{center}

\vspace{0.5cm}
\hrule \vspace{0.1cm}
\vspace{0.2cm}
{ \noindent \textbf{Abstract}
\vspace{0.3cm}

We initiate a study of subregion dualities, entropy, and redundant encoding of bulk points in holographic theories deformed by $T\bar T$ and its generalizations. This includes both cut off versions of Anti de Sitter spacetime, as well as the generalization to bulk de Sitter spacetime, for which we introduce two additional examples capturing different patches of the bulk and incorporating the second branch of the square root dressed energy formula.
We provide new calculations of entanglement entropy (EE) for more general divisions of the system than the symmetric ones previously available. We find precise agreement between the gravity side and deformed-CFT side results to all orders in the deformation parameter at large central charge. An analysis of the fate of strong subadditivity for relatively boosted regions indicates nonlocality reminiscent of string theory. We introduce the structure of operator algebras in these systems. The causal and entanglement wedges generalize to appropriate deformed theories but exhibit qualitatively new behaviors, e.g. the causal wedge may exceed the entanglement wedge. This leads to subtleties which we express in terms of the Hamiltonian and modular Hamiltonian evolution.   Finally, we exhibit redundant encoding of bulk points, including the cosmological case.  

\vspace{0.4cm}

 \hrule

\vspace{0.6cm}}
\end{titlepage}

\tableofcontents

\section{Introduction}\label{sec:intro}

In recent years, holographic dualities have developed in several important ways.  In AdS/CFT, the  
association of bulk regions with appropriate operator algebras in the dual `boundary' theory leads to an in-principle method for their approximate reconstruction \cite{Lewkowycz:2013nqa,Faulkner:2013ana,Jafferis:2015del,Dong:2016eik,Faulkner:2017vdd}, moving beyond the original HKLL prescription developed earlier in \cite{Hamilton:2005ju}.\footnote{See \cite{Harlow:2018fse}\ for a review.}  This in turn leads to a lesson that the encoding of a bulk point in the dual is redundant, as in quantum error correction \cite{Almheiri:2014lwa}.

Although it is an extraordinarily fruitful case study for quantum gravity, AdS/CFT is neither phenomenologically viable nor generic in string theory, with its special asymptotic boundary and bulk geometry being highly unrealistic.
In another line of development, the $T\bar T$ deformation
\cite{Zamolodchikov:2004ce, Smirnov:2016lqw,Cavaglia:2016oda, Dubovsky:2012wk}\ and its generalizations such as \cite{Giveon:2017nie, Hartman:2018tkw,Gorbenko:2018oov, Jafari:2019qns}
enable us to isolate a finite patch of spacetime not intersecting the original boundary \cite{McGough:2016lol}.  This corresponds to a Dirichlet boundary condition for the metric, and also for additional bulk fields given the prescription \cite{Hartman:2018tkw}.  (A related deformation which accounts for bulk matter intrinsically is the single-trace version developed in \cite{Giveon:2017nie}.) 

Meanwhile, holographic descriptions of the realistic case of bulk de Sitter geometry  \cite{Strominger:2001pn, Strominger:2001gp,Alishahiha:2004md,Alishahiha:2005dj,Dong:2010pm,Dong:2011uf,Anninos:2011ui,Anninos:2012qw,
Anninos:2017hhn,Dong:2018cuv,Gorbenko:2018oov,Gross:2019ach, Cotler:2019nbi, Cotler:2019dcj}\ have developed significantly.  In particular, patches of de Sitter spacetime, including the dS/dS patch covering more than an observer region, arise from appropriate generalizations of the $T\bar T$ deformation \cite{Zamolodchikov:2004ce, Smirnov:2016lqw,Cavaglia:2016oda, Dubovsky:2012wk}\ as explained recently in \cite{Gorbenko:2018oov,Gross:2019ach}. These are described by a trace flow equation of the form
\be\label{eq:introflowQFTvar}
\tr \,T = -\frac{c}{24\pi} {\cal R}^{(2)} - 4\pi \lambda T\bar T +\frac{c_2}{\pi\lambda}\,,
\ee   
with $c$ the central charge, $\mc R^{(2)}$ the scalar curvature of spacetime, and $c_2$ a constant. Additional bulk matter fields with Dirichlet boundary conditions lead to extra terms in the trace flow equation~\cite{Hartman:2018tkw}, but there are still many interesting observables where such terms are not excited or are subleading (as will be our case).

In setting up the present work in \S\ref{sec:setup}, we will provide two new examples of this.  One is a corollary of \cite{Gorbenko:2018oov}\ which doubles the space of solvable and universal \cite{Zamolodchikov:2004ce, Smirnov:2016lqw, Cavaglia:2016oda}\ deformations of 2d quantum field theories. When interpreted holographically, this formulates the static patch of $\text{dS}_3$ at the level of pure gravity.  The other is an extension of the trajectory defined in \cite{Gorbenko:2018oov}\ which connects to another branch of a square root appearing in the formula for the energy levels; this formulates the full dS/dS patch of de Sitter spacetime with one extended trajectory.  

These $T\bar T+\dots$ prescriptions for radially bounded patches of bulk (A)dS spacetime can be viewed as another form of subregion duality.  
It is natural to combine the two notions of subregion, and investigate the extension of reconstructions in \cite{Hamilton:2005ju,Lewkowycz:2013nqa,Faulkner:2013ana,Jafferis:2015del,Dong:2016eik,Faulkner:2017vdd,Harlow:2018fse} 
to the more general case of CFTs deformed by $T\bar T$ and its generalizations, including the realistic cosmological case. This is directly related to the behavior of the entanglement entropy in such deformed theories, something that we will study in detail in this work using both sides of the duality.

In order to carry this out, we must determine the effect of the deformation on the causal and entanglement wedges defined e.g. in  \cite{Headrick:2014cta}, their associated algebras, and the action of the Hamiltonian and modular Hamiltonian.  The $T\bar T+\dots$ deformations do not produce local quantum field theories, and a priori one must not take for granted properties like causality and locality of the operator algebras. We will find several specific manifestations of the nonlocality, which enables novel relations between the causal wedge (CW) and the entanglement wedge (EW).  
In the case of the causal wedge, we find a subset of deformed theories (specific examples being the dS/dS theory \cite{Gorbenko:2018oov}\ and the cutoff version of Poincar\'e AdS) for which the original notion persists in the semiclassical bulk theory because boundary to boundary signals travel subluminally and fastest along the boundary.\footnote{The stability of Dirichlet cutoffs in semiclassical general relativity is a subject of active investigation.  It would be interesting to generalize the analyses of e.g. \cite{Andrade:2015gja}\ to the full range of bulk/boundary geometries we consider here.  As observed in \cite{Gorbenko:2018oov}, the superluminality in \cite{Marolf:2012dr}\ does not persist in the boundary dS cases.}  In these cases, HKLL \cite{Hamilton:2005ju}\ applies to our case, and we note the appearance of a causal shadow which somewhat limits the reconstructions.  Once we include the prescription \cite{Hartman:2018tkw}, we note that the operators are local on the Dirichlet wall.  In the asymptotic AdS case, HKLL and other bulk reconstruction prescriptions become more complex as one proceeds inward in the bulk.  In the present context, having deformed the CFT via $T\bar T+\dots$, the operators start essentially local on the finite Dirichlet wall.  HKLL then starts to render them nonlocal as we move inward from that locus.  In essence, the complication that arose in AdS/CFT at the radial position of the cutoff surface is replaced by the nontrivial deformation of the theory itself; although the deformed theory contains nonlocal features, there is emergent bulk locality down to the bulk string scale even in the presence of the Dirichlet wall.  
In the case of the entanglement wedge, we specify a division of the system which semiclassically  corresponds to the division across the extremal surface of \cite{Ryu:2006bv, Hubeny:2007xt}\ as in \cite{Lewkowycz:2013nqa, Faulkner:2013ana}.

\subsection{Summary of results}\label{subsec:summary}

Let us now describe our main results.  We analyze in \S\ref{sec:bSSA} and \S\ref{sec:subregion} the R\'enyi and Von Neumann entropies on both sides of the duality in two case studies, generalizing the method of \cite{Donnelly:2018bef}\ to less symmetric divisions of the system. This reveals two striking properties of $T \bar T$. First, we find that all contributions of the deformation to the entropy turn out to localize at the endpoints of the entangling region
\be
L S'(L) = 2 \times \lim_{n\to 1} (2\pi n) \int_0^{\rho_0 \ll L} \rho d\rho\, n \partial_n \langle \tr\,T(\rho)\rangle\,.
\ee
Here $L$ is the size of the interval for which we compute the EE, $n$ is the replica index, and $\rho$ is the radial distance to one of the endpoints. A similar expression is valid in the dS case, with $L$ replaced by the curvature scale. Evaluating this requires then calculating the change in the stress tensor under a change in the replica opening angle, $n \partial_n \langle \tr\,T \rangle|_{n=1}$, near the endpoints. We obtain this by solving the trace flow and conservation equations,
\be\label{eq:introTCform}
n \partial_n \langle \tr\, T (\rho)\rangle |_{n \to 1}=\epsilon\, \frac{c}{3}\,\frac{\lambda c}{24\pi}\,\frac{C(\frac{\lambda}{L^2})^2}{ \rho^4\left(1+\epsilon \frac{\lambda c}{6\rho^2} C(\frac{\lambda}{L^2}) \right)^{3/2}}\,,\; \epsilon \equiv 1-n\,,
\ee
with $C(\lambda/L^2)$ a constant that we discuss shortly. Rotational symmetry is restored at the tips of the replica manifold, providing a crucial simplification that is at the root of our exact results.

This, and related expressions we present for other components of the stress tensor, exhibit the second feature we find about $T \bar T$, namely that the deformation smooths out the singularities from the conical defects at the endpoints. This is reflected in the nonperturbative shift in the denominator, controlled by $\epsilon \lambda$.  Combining these two equations gives that $L S'(L) = \frac{c}{3} C(\lambda/L^2)$. So the function $C(\lambda/L^2)$ encodes the behavior of the entanglement entropy as well as the twist operators in the deformed theory. We show that a similar result holds in the de Sitter thermal calculation.

These features allow us to generalize the CHM map~\cite{Casini:2011kv}, originally envisioned for CFTs, to $T \bar T$ deformed CFTs. This maps the domain of the dependence of an interval in Poincar\'e space to the static patch of de Sitter. Due to the localization property of $T \bar T$, the entanglement entropy for the interval becomes the same as the thermal entropy in de Sitter. This allows us to derive an expression for $C(\lambda/L^2)$ and for the entropy $S(L)$ for an interval of size $L$, to all orders in the deformation:
\be\label{eq:EEintervalintro}
L S'(L)= \frac{c}{3}\,\frac{1}{\sqrt{1+ \frac{\lambda c}{3L^2}}}\,,
\ee
with $\lambda$ the strength of the $T \bar T$ deformation;
see \S\ref{subsec:defCFT} for more details. This matches exactly the holographic answer, and provides another instance of an exact calculation in the presence of $T \bar T$ beyond e.g. the energy level formula~\cite{Smirnov:2016lqw} (albeit here we need to use large $c$). 

In this way, we establish the Ryu-Takayanagi formula for a single interval in the radially cutoff AdS Poincar\'e patch.  This lends support to the possibility that the general proof \cite{Lewkowycz:2013nqa}\ for AdS/CFT may extend to our deformed theory, something that will be interesting to nail down in the future. See also~\cite{Murdia:2019fax} for work in this direction.

The interval entropy (\ref{eq:EEintervalintro}) violates boosted strong subadditivity \cite{Casini:2012ei}, indicating that additional operators join the algebra under a relative boost of subregions.  This is consistent with causality and helps to characterize the non-locality of the theory. 
In the earlier work \cite{Chakraborty:2018kpr}, a contribution to the von Neumann entropy was also found at first order in the single-trace version of the deformation; they were working with the opposite sign of the deformation from ours, the sign that leads to a Hagedorn spectrum as opposed to our case of interest here with a finite entropy.\footnote{The recent works \cite{Chen:2018eqk, Jeong:2019ylz}\ also studied entanglement entropy in $T\bar T$ deformed theories, although they did not find this first order effect.} 

The second example we analyze in detail in \S\ref{sec:subregion} is the deformed theory dual to a dS/dS warped throat.\footnote{Other recent works that studied the EE for $T \bar T$ on de Sitter include~\cite{Geng:2019bnn, Grieninger:2019zts}.} Here the deformation, recently introduced in~\cite{Gorbenko:2018oov}, is defined by a coordinated flow (\ref{eq:introflowQFTvar}) that includes $T \bar T$ and a 2d cosmological constant.
For a a subsystem that is half of the space, we evaluate the partition function for an $n$-sheeted cover of the sphere,
\be\label{eq:Znintro}
\partial_r\,\log Z_n= -2 \pi n r\,\int_0^\pi d\theta\,\sin\theta\,\,\langle \tr T \rangle\,.
\ee
We then show that the $0$-th R\'enyi entropy (the log of the dimension of the reduced Hilbert space) agrees with the entanglement entropy,
\be
S_0(r) = S_1(r)=\frac{\pi c}{6} \;\;\text{for}\;\;r=\sqrt{\frac{c\lambda}{12 }}\,.
\ee
This implies that the state for the subsystem is maximally mixed. In the holographic side, the value of $r$ above corresponds to the central slice of $dS_3/dS_2$,  $w_c= \frac{\pi}{2} \ell$. Given this result, we determine that states associated to subsystems with size different than half the space behave as random pure states.

The combination of entanglement and causal wedges introduces new features in the deformed theories as compared to asymptotic AdS/CFT, which are analyzed in \S\ref{sec:subregion}. In particular, the causal wedge of a region $R$ can exceed its entanglement wedge, and can overlap with the entanglement wedge of the complementary region $\bar R$.  This implies a novel commutator structure of the associated algebras, which we describe. We argue that in this case the modular evolution in $R$ does not commute with the time evolution in the causal domain of the complement: 
\be\label{eq:novel2intro}
[\rho_{\bar R}, U_{D[ R]}]\ne 0 \,.
\ee
These features of the causal wedge and entanglement wedge algebras also explain the violation of the boosted SSA discussed above.   

Finally, having characterized the subregions we return in \S\ref{sec:QEC} to one of the motivating questions:  does the redundancy of bulk point encoding (a.k.a. quantum error correction) \cite{Almheiri:2014lwa}\ survive these deformations, in particular the extended trajectory \cite{Gorbenko:2018oov}\ that is required for the cosmological case?  We find indeed that redundant encoding continues to occur, and we indicate some requirements for toy models of this effect that might generalize the tensor network toy examples in asymptotic AdS/CFT, which might be used for near term simulations for quantum cosmology. 


\section{Setup:  (A)dS patches and $T\bar T+\dots$ trajectories}\label{sec:setup}

\begin{figure}[h!]
\begin{center}  
\includegraphics[width=.9\textwidth]{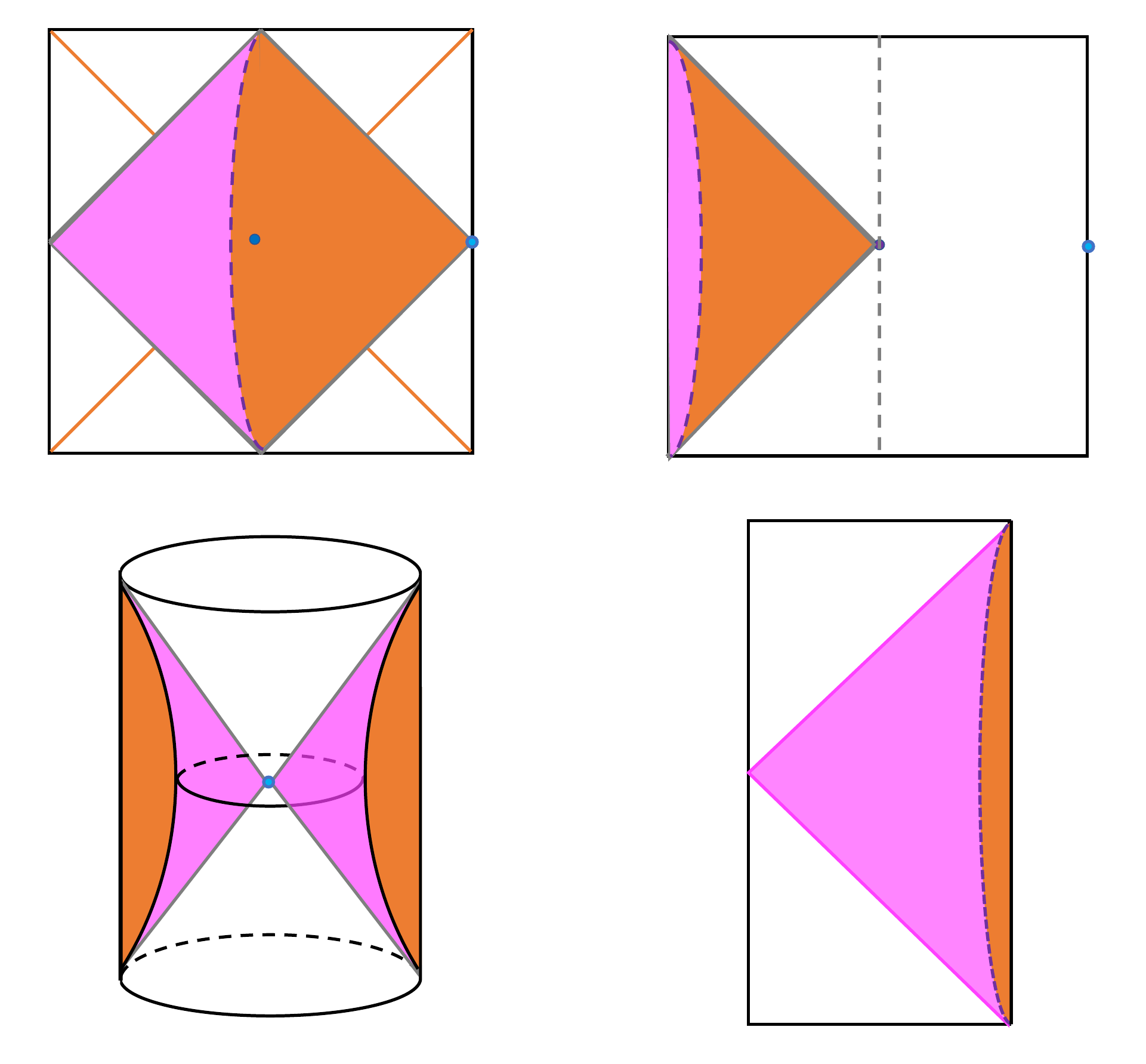}
\captionsetup{width=0.9\textwidth}
\caption{Patches we will work with depicted in purple within the $\text{AdS}$ and $\text{dS}$ Penrose diagrams.  The top left is $\text{dS}/\text{dS}$, with a fixed $w=w_c$ slice indicated by the dashed line.  The purple and orange together cover the full dS/dS patch of $dS_3$, while the purple indicates the region that remains after introducing the cutoff at a fixed $w_c$ in the coordinate system (\ref{mets}).    The top right is $\text{dS}/{\text{cylinder}}$, with again a fixed radial position $r=r_c$ indicated by the dashed line.  The bottom left similarly depicts cut off $\text{AdS}/\text{dS}$ and the bottom right (a slice of) cut off $\text{AdS}/{\text{Poincar\'e}}$.}
\label{fig:patches}
\end{center}  
\end{figure}

We are interested in the the holographic formulation of various finite patches of AdS and dS spacetime, obtained via the $T\bar T$ deformation \cite{Zamolodchikov:2004ce, Smirnov:2016lqw,Cavaglia:2016oda, Dubovsky:2012wk} and some of its recent generalizations \cite{Gorbenko:2018oov, Hartman:2018tkw}.
For simplicity, we focus on three bulk and two boundary dimensions (along with appropriate compact dimensions that arise internally in string theory), although very interesting generalizations to other dimensions are available in \cite{Hartman:2018tkw,Gross:2019ach}.
The 3d bulk case is the lowest dimensionality in which putative spatial boundary subregions exist, and its 2d dual makes use of all the methods available in the original works on $T\bar T$.   
We will consider cases where the bulk theory (in its vacuum) is either AdS or dS, with a Dirichlet boundary that is either flat or de Sitter.  These varieties of bulk/boundary will be denoted
AdS/Poincar\'e, (A)dS/cylinder, and (A)dS/dS.  See Fig.~\ref{fig:patches} for a depiction of the patches we will consider within the Penrose diagrams of AdS and dS. 

The warped metrics for each case in the vacuum state are as follows.
\bea\label{mets}
ds^2_{\text{(A)dS/dS}} &=& dw^2+\sin(\text{h})^2\frac{w}{\ell}\left(-d\tau^2+\ell^2\cosh^2\frac{\tau}{\ell} d\phi^2\right) ~~~ w\le w_c \nonumber\\
ds^2_{\text{(A)dS/cylinder}}&=&-(1\pm r^2/\ell^2)dt^2 +  \frac{dr^2}{1\pm r^2/\ell^2}+r^2 d\phi^2~~~ r\le r_c \nonumber\\
ds^2_{\text{AdS/Poincar\'e}} &=&\ell^2 \frac{-dt^2+dx^2+dz^2}{z^2} ~~~z\ge z_c\nonumber\\
\eea
We note that in the dS/cylinder case, and in the version where we cover the full dS/dS region with one bounded patch at $w_c=\pi\ell$, the boundary is in the infrared (most gravitationally redshifted region), something that is far from the situation in AdS/CFT.  In the other cases, AdS/cylinder, AdS/Poincar\'e, and AdS/dS, and dS/dS with $w_c\le \pi\ell/2$, the boundary is at the most UV slice of the geometry.  In the (A)dS/dS and AdS/Poincar\'e cases, there is another important feature:  signals travel fastest along the boundary. This is reminiscent of the feature identified in \cite{Gorbenko:2018oov}\ that the boundary gravitons are luminal rather than superluminal in this case.       
We will find that these distinctions are significant in our studies of subregion dualities, with the examples that are most similar to AdS/CFT being the most amenable to redundant encodings (error correction).  But they all admit a formulation in terms of specific trajectories including and generalizing $T\bar T$.     

To begin, we will review and extend the formulation of the deformed CFTs of interest in a unified way, introducing two new examples beyond those explicitly covered in the existing references. These are the static patch of de Sitter,\footnote{The static patch of $\text{dS}_2$ appeared also in the interesting recent work \cite{Gross:2019ach}\ which provides a tractable formulation of a 1d analogue of $T\bar T$ and its generalizations such as \cite{Gorbenko:2018oov}\ in terms of a dual quantum mechanics theory, with connections to \cite{Anninos:2017hhn}.}  and the dS/dS patch obtained via a single extended trajectory rather than via a joining of two warped throats.  As we will see shortly, the dS static patch (a.k.a. dS/cylinder) has the virtue that its pure gravity dual, a corollary of the deformation derived in \cite{Gorbenko:2018oov}, is as universal and solvable even at finite $c$ as the original $T\bar T$ deformation, via the methods introduced in \cite{Zamolodchikov:2004ce, Smirnov:2016lqw}.  Regardless of holographic duality, this doubling of the space of such calculable deformations may be of interest in its own right in the study of 2d solvable models.

We will work with an integrated deformation by the irrelevant operator ``$T\bar T$''
\be\label{TTbardef}
T\bar T \equiv \frac{1}{8}(T_{ab}T^{ab}-(T^a_a)^2)\,.
\ee
In some situations, this operator factorizes.  This is true in all our examples
at least at large $c$, along with $1/c$ corrections that can systematically be included.\footnote{The works~\cite{Tolley:2019nmm, Mazenc:2019cfg} provide a definition of the $T \bar T$ deformation in curved space, which could be used to study finite $c$ effects.}
For the cylinder, the factorization occurs for all $c$ \cite{Zamolodchikov:2004ce}.  The trajectories defining the deformed CFT can be characterized at the level of pure 3d gravity by 
a differential equation for the log of the partition function:
\be\label{eq:ZTTbLambda}
\frac{\partial}{\partial \lambda}\, \log Z =-2\pi \,\int d^2 x\,\sqrt{g}\, \langle T \bar T \rangle+\frac{1-\eta}{2\pi \lambda^2}\int d^2 x \,\sqrt{g}
\ee
Here $\eta=1$ corresponds to the initial trajectory starting from the seed CFT at $\lambda=0$, with holographic dual a patch of
bulk AdS$_3$.  Once we are along this trajectory, at some nonzero $\lambda$, we can join onto a trajectory with $\eta\ne 1$.
As explained in detail in \cite{Gorbenko:2018oov}, such an extension of the trajectory to one with $\eta=-1$ is appropriate for bulk dS$_3$ (and $\eta=0$ for a flat bulk spacetime).  For many purposes we can formulate the trajectory via the
the trace flow equation
\be\label{flowQFTvar}
T^a_a = -\frac{c}{24\pi} {\cal R}^{(2)} - 4\pi \lambda T\bar T -\frac{{\eta}-1}{\pi\lambda}\,,
\ee
where $T_{ab}$ is the stress energy tensor of the $2d$ theory, which satisfies the conservation equations 
\be\label{Tcons}
\nabla^a T_{ab}=0\,.
\ee
The various cases of interest are as follows:
\bea\label{TrajDuals}
\text{AdS/dS} &:& \eta=1, ~~~~0<\frac{c}{24\pi} {\cal R}^{(2)}<\infty \nonumber\\
\text{dS/dS} &:& \eta=-1, ~~~~  \frac{1}{\pi\lambda}<\frac{c}{24\pi} {\cal R}^{(2)}<\infty \nonumber\\
\text{AdS/cylinder} &:& \eta=1, ~~~~ \frac{c}{24\pi} {\cal R}^{(2)}=0 \nonumber\\
\text{dS/cylinder}   &:& \eta=-1, ~~~~\frac{c}{24\pi} {\cal R}^{(2)}=0\,,
\eea
and again we note that in the latter two cases, factorization of the $T\bar T$ operator is valid at finite $c$ via the derivation in \cite{Zamolodchikov:2004ce}.  Below we will describe two versions of the dS/dS case.  The original ultimately involves a joined system of two warped throats, each cut off by a Dirichlet wall at $w_c=\pi\ell/2$ and formulated by its own trajectory (\ref{eq:ZTTbLambda}-\ref{flowQFTvar} ) as described in \cite{Gorbenko:2018oov}.  Another option, as we will see shortly, involves a single extended trajectory to obtain the full dS/dS patch bounded by the slice $w_c=\pi\ell$.

There is detailed evidence from calculations of energies and entropies supporting the conjectured holographic dualities between a Dirichlet wall-bounded patch of gravity with cosmological constant $\Lambda_3=-2\eta/\ell^2$ and the deformed-CFT trajectories.  We summarize this and extend it to our new examples in the next two subsections.    

\subsection{Dressed Energies and additional dualities}

For the cylinder (or Poincar\'e) cases where the 2d curvature $R=0$, the $T\bar T$ operator factorizes as in \cite{Zamolodchikov:2004ce}\ and for the full two dimensional space of couplings parameterized by $\lambda$ and $\eta$ we can calculate the energy spectrum exactly at finite $c$. This gives
\be\label{Ecyl}
E=-2\pi L T^t_t=\frac{L}{\pi\lambda}\left(1-\sqrt{\eta-\frac{4\pi^2\lambda(\Delta+\bar\Delta-c/12)}{L^2}+\frac{4\pi^4\lambda^2(\Delta-\bar\Delta)^2}{L^4}}\right)
\ee
Here $L=2\pi r$ is the spatial size of the cylinder on which the 2d theory lives, and $\Delta, \bar\Delta$ are the left and right moving dimensions of the state in the seed theory, which we have taken to be a 2d CFT.  This formula agrees with the quasilocal energy of the corresponding patch of spacetime in a theory with bulk cosmological constant $\Lambda_3=-2\eta/\ell^2$ spacetime \cite{McGough:2016lol, Gorbenko:2018oov}\ of either sign.

In the (A)dS/dS cases, the differential equation for the dressed energy similarly leads to a solution of the form
\be\label{dressedLsolns}
\langle T^\tau_\tau \rangle=\frac{1}{\pi\lambda}\left(1\mp\sqrt{\eta+c \frac{{{\cal R}^{(2)}\lambda}}{24}- \frac{C_1\lambda}{L^2}} \right)\,.
\ee
In this curved case (and for any boundary geometry with bulk matter excitations), one requires use of large $c$ factorization in typical states on the deformed-QFT side.  This corresponds to semiclassical gravity in a finite patch of spacetime, suggesting that it can in principle be supplemented by perturbative corrections in $1/c$ using UV-finite perturbative string corrections.  At finite $c$, there may be ambiguities or fundamental limitations on this definition of the theory.  Indeed, in string theory de Sitter is only metastable, so its more complete formulation likely requires its decaying phase into a more general FRW solution, something that admits an analogous description in terms of two coupled sectors \cite{Dong:2011uf}.  In the present work, we will focus on the exponentially long lived de Sitter phase although we expect some of the phenomena we derive to extend to the later FRW phase.\footnote{Another approach to extending a dS patch to a completely formulated system is analyzed in \cite{Anninos:2017hhn}.}

The top sign in (\ref{dressedLsolns}), with $C_1=0$, reproduces the quasilocal energy of one of the two warped throats of the dS/dS patch, with $w_c\le \pi\ell/2$ corresponding to ${\cal R}^{(2)}\ge 24/\lambda c$ \cite{Gorbenko:2018oov}.   At that limiting value, the square root vanishes; on the gravity side this corresponds to the vanishing extrinsic curvature in the central slice of the dS/dS patch of dS$_3$.  In the full dS/dS correspondence, we construct two such warped throats, and join them on a common UV slice by integrating over their shared metric, leading to a flat entanglement spectrum \cite{Dong:2018cuv}.    

There is another interesting option at this point, however, which brings in a role for the other branch of the square root in the energy formula (\ref{dressedLsolns}).  In the original $T\bar T$ deformation \cite{Smirnov:2016lqw, Cavaglia:2016oda}, the top sign was unambiguously chosen in order to match smoothly to the seed QFT in the $\lambda\to 0$ limit.  We inherit this sign as well in our extended trajectory building up a dS/dS throat as in \cite{Gorbenko:2018oov}.  But once we reach the end of that trajectory where the square root vanishes, we may smoothly continue through to the other sign of the square root.  It is a simple exercise to check that this reproduces the quasilocal energy of a larger portion of the dS/dS patch; we can proceed all the way to $w_c=\pi\ell$ in this way.  

This last example, like the static patch example, has the property that the boundary is then at one of the most infrared (highly redshifted) slices of the warped geometry.  These cases, which in this respect are farther from AdS/CFT than the other versions of the duality, will exhibit less optimal features in terms of subregion dualities.  Nonetheless they provide new examples of $T\bar T$ generalizations with interesting features and with holographic interpretations.

\subsection{Stress energy and entropy}\label{subsec:Tentropy}

It is also interesting to study the density matrix and entropies associated with various divisions of the system.  
In the (A)dS/dS case this has led to another test of the duality obtained via calculations on both sides of the Von Neumann and R\'enyi entropies for a particularly simple division of the system into halves.  This was pioneered in  \cite{Donnelly:2018bef}\ for the AdS/dS case and straightforwardly generalized to dS/dS in \cite{Gorbenko:2018oov}.

One of our main technical points in the present work will be to generalize the entropy calculations to more generic divisions of the system as well as extending the calculations to capture essential properties of the density matrix (or equivalently its log, the modular Hamiltonian) itself.  

This is interesting in itself in the deformed CFT as a way of probing its novel properties, independently of holography.  For holography, this will enter into our analysis of the fate of subregion dualities and the relations between bulk and boundary modular flow \cite{Jafferis:2015del, Faulkner:2013ana, Faulkner:2017vdd}.  

It is not generically easy to calculate entanglement entropy in an interacting theory.  But the original calculations of the dressed energies illustrate the special tractability of the $T\bar T$ deformation and its relatives, and it is reasonable to explore to what extent that extends to calculations of other physical quantities.  
Indeed the equations governing the dressed stress energy $T_{ab}$ provide a method to extract entanglement entropy and properties of the modular flow in some cases.  This was introduced and illustrated in a particular, symmetric example in \cite{Donnelly:2018bef}.
One result of the present work will be to extend this to new, less symmetric, examples.

At large $c$, $T_{ab}$ is determined by the trace flow equation and stress-energy conservation
\bea\label{Teqs}
\nabla_a T^a_b &=& 0 \\
T^a_a &=& -\frac{ c {\mc R^{(2)}}}{24\pi}+\frac{c_2}{\pi\lambda}-\frac{\pi\lambda}{2}(T^{ab}T_{ab}-(T^a_a)^2), ~~~~ c_2=1-\eta \nonumber
\eea  
with appropriate boundary conditions.  
In general, these form a quasilinear system of two partial differential equations (PDEs); as we will see, these sometimes admit a solution via the method of characteristics.  In appendix \ref{sec:Charappendix}\ and \S\ref{sec:dScharspacman}\ we will investigate the characteristics and apply them to our problem.  


We can apply the solutions for $T_{ab}$ in two ways to study the physics of the reduced density matrix $\rho$ appropriate to a given division of the system.  First, as in \cite{Donnelly:2018bef}, if we can solve these equations for on the $n$-sheeted replicated geometry arising in the path integral calculation of $\tr \rho^n$. Denoting the replicated partition function by $Z_n$, the (modified) R\'enyi entropy is defined as
\be\label{eq:St-def}
\tilde S_n(L)= (1-n \partial_n)\,\log Z_n\,,
\ee
where $L$ is the overall scale of the system.\footnote{For concreteness, in this paper we would think of $L$ as the size of the interval in flat space in our AdS/Poincar\'e case study, or the Euclidean de Sitter (sphere) radius for an interval in (A)dS/dS.}
The von Neumann entropy arises as
\be\label{eq:St-S}
S(L) = \lim_{n\to 1} \tilde S_n(L)\,.
\ee
We can then use the relation
\be\label{eq:ZnTrTSn}
L \frac{d}{dL}\,\log Z_n = - \int d^2x\,\sqrt{g}\,\langle\text{tr}\,T \rangle \;,\; L \frac{d}{dL}\, \tilde S_n = - (1-n \partial_n)\,\int d^2x\,\sqrt{g}\,\langle\text{tr}\,T \rangle \,.
\ee
to obtain the corresponding entropy. We will illustrate in \S\ref{sec:bSSA} how the trace flow and conservation equations lead to an exact large $c$ result for the entanglement entropy in a finite interval of length $L$ in Minkowski space. 
Then in \S\ref{sec:wcpiover2}\ we will illustrate this for a dS/dS case study. These provide interesting instances where a nontrivial partition function can be evaluated to all orders in the deformation, something that, as we stressed already in \S \ref{sec:intro}, is a consequence of the special properties of $T \bar T$.

A second application of these equations pertains to the behavior of the density matrix itself, $\rho=e^{-K}$ with $K$ the modular Hamiltonian.      
In a standard local theory of quantum fields $\phi$, with a division of the system into a spatial region $R$ and its complement, an entry $\rho_{R, \phi^-(x),\phi^+(x)}$ in the density matrix (with $x\in r$) is computed by the following path integral.  Starting from the Euclidean path integral that constructs the partition function, we cut it open on the region $R$ and impose boundary conditions $\phi^\pm$ on the top and bottom of the cut.
We will refer to this cut geometry as the pac man.

In our nonlocal 2d theories, we cannot generically assume a precise division of the system into spatial subregions.  Still, for the theories which are holographically dual to an emergent semiclassical bulk gravitational spacetime patch, the bulk effective theory is local down to the string scale, and one can divide the system across an extremal surface as in the familiar AdS/CFT context.  This defines the density matrix via a semiclassical gravitational path integral.  
The resulting prediction for the deformed-QFT side is a modification of the action of $\rho$ (equivalently $K$) as an operator.  In \S\ref{sec:wcpiover2}\ we will analyze this directly on the 2d deformed-CFT side in a special case of interest (dS/dS with $w_c=\pi\ell/2$) using the behavior of the stress energy derived from the basic equations (\ref{Teqs}).   

\subsection{Revisiting the entropy calculation in AdS/dS}\label{subsec:DS}

Before proceeding to our new results, we would like to revisit the calculation in~\cite{Donnelly:2018bef} from a different point of view, which will be useful in the following sections. This work computed the entanglement entropy for a deformed CFT on $\text{dS}_2$, when the spatial region is half of the full system. This corresponds to the thermal entropy for the de Sitter static patch. The time translation symmetry  allows one to obtain the entropy at large $c$ and to all orders in (fixed) $\lambda c$, and the result matches the holographic answer for cutoff AdS sliced by dS.  This calculation also generalized readily to the dS/dS case \cite{Gorbenko:2018oov}. 

We will now establish a very special feature of the $\lambda T \bar T$ deformation, namely that all the dependence of the entanglement entropy on $\lambda$ comes from the endpoints of the entangling region. This will turn out to be closely related to fact that the deformation smooths out the conical singularities introduced for the R\'enyi entropies. 

Let us perform the Euclidean calculation of the entropy. The metric is
\be\label{methem}
ds^2=r^2(d{\theta}^2 + \sin^2{\theta} d{\phi}^2)\,,
\ee
with $\phi$ the analytic continuation of the $dS_2$ static patch time, and $r$ the dS radius. The system is divided in half, $R=\pi\ell$, and the region $R$ is the locus $\phi=0, 0\le\theta\le\pi$. 
For this metric,
using the Christoffel symbols
\be\label{Christoff}
\Gamma^\theta_{\phi\phi}=-\cos\theta\sin\theta, ~~~~ \Gamma^\theta_{\phi\theta}=\cot\theta
\ee
we find that the equations (\ref{Teqs}) become 
\bea\label{TeqsS}
0 &=& \partial_\phi T^\phi_\phi+\partial_\theta(\sin^2\theta T^\phi_\theta)+  T^\phi_\theta \cos\theta\sin\theta \nonumber\\
0 &=& \partial_\theta T^\theta_\theta+\partial_\phi(T^\phi_\theta)+  (T^{\theta}_{\theta}-T^{\phi}_{\phi} )\cot{\theta} \nonumber\\
T^{\theta}_{\theta}+T^{\phi}_{\phi} &=& -\frac{ c {\mc R^{(2)}}}{24\pi}+\pi\lambda(T^{\theta}_{\theta} T^{\phi}_{\phi}-(T^{\phi}_{\theta})^2\sin^2{\theta}) \,.
\eea 
In fact we need to work with the smoothed out replicated geometry,  taking (inspired by~\cite{Fursaev:1995ef})
\be
ds^2=r^2 \left(\frac{\sin^2\theta+ n^2 \delta^2}{\sin^2\theta+\delta^2} d\theta^2+n^2 \sin^2\theta\,d\phi^2 \right)\,.
\ee
As the regulator $\delta \to 0$, we recover the replicated manifold. This is an Einstein space, with curvature
\be
{\cal R}^{(2)}=\frac{2}{r^2} \left(1- \delta^2 \frac{(n^2-1)(1+n^2 \delta^2)}{(\sin^2\theta+ \delta^2 n^2)^2} \right) \approx \frac{2}{r^2} \left(1- \delta^2 \frac{n^2-1}{(\sin^2\theta+ \delta^2 n^2)^2} \right)  \,.
\ee
In particular, near the smoothed-out ends of the interval $R$,
\be
{\cal R}^{(2)} =  \frac{2}{r^2} \frac{1-n^2}{\delta^2 n^4} + O(\delta^0)\,,
\ee
so we have positive curvature for $n \le 1$ and negative curvature for $n \ge 1$.  For some of our applications, we will be interested in the von Neumann entropy, and hence take $n\to 1$; for others we will keep $n$ finite and analyze the R\'enyi entropies in their own right.  

\subsubsection*{Energy-momentum tensor}

In order to compute R\'enyi entropies using the relation (\ref{eq:ZnTrTSn}),
the first step is to compute the vacuum expectation values of the stress tensor. The space is Einstein but not maximally symmetric, so we do not have $T_{ab} \propto g_{ab}$ (due to the $\theta$ dependence in the curvature). 
We do still have a simplification in the equations (\ref{TeqsS}) from the symmetry in the $\phi$ direction, which enables us to set $T^\phi_\theta=0$ in the vacuum and seek a $\phi$-independent solution for the other components.   

In terms of the variables
\be
U \equiv \pi \lambda T^\theta_\theta-1\;,\;V\equiv \pi \lambda T^\phi_\phi-1\,,
\ee
one finds as in \cite{Donnelly:2018bef}\ that the conservation equations and the flow equation read
\bea
&&UV = \frac{c\lambda}{24} {\mc R}^{(2)}(\theta) + \eta \nonumber\\
&& \partial_\theta U + \cot \theta (U-V) =0\,.
\eea
For generality we have kept the parameter $\eta$ in (\ref{Teqs}); for the present discussion we have $\eta=1$.
The solution that is nonsingular at the tips and gives the right branch is
\be
U = - \sqrt{\eta + \frac{c\lambda}{12 r^2} \left(1+ \frac{(1-n^2)(1+n^2\delta^2)}{n^2(\sin^2\theta+n^2 \delta^2)} \right)}\,.
\ee
This corresponds to the stress energy components
\bea\label{eq:Tvevs}
T^\theta_\theta&=& \frac{1}{\pi \lambda} \left(1- \sqrt{\eta + \frac{c\lambda}{12 r^2} \left[ 1+ \left(\frac{1}{n^2}-1 \right) \frac{1+n^2\delta^2}{\sin^2\theta + n^2 \delta^2}\right] } \right) \\
T^\phi_\phi &=&\frac{1}{\pi \lambda} \left(1- \frac{\eta+ \frac{c \lambda}{12 r^2} \left[1-\delta^2 \frac{(n^2-1)(1+n^2\delta^2)}{(\sin^2\theta+ \delta^2 n^2)^2} \right]      }{\sqrt{\eta + \frac{c\lambda}{12 r^2} \left[ 1+ \left(\frac{1}{n^2}-1 \right) \frac{1+n^2\delta^2}{\sin^2\theta + n^2 \delta^2}\right] } }   \right) \nonumber\,.
\eea
This generalizes the expressions in~\cite{Donnelly:2018bef} by including the effect of $\eta$ as in \cite{Gorbenko:2018oov}, and the smoothing parameter $\delta$ which we have treated slightly differently, but consistently with the earlier results. For $n>1$, as noted in \cite{Donnelly:2018bef}\ we find complex energy levels, unless
\be
\left(\eta+ \frac{c\lambda}{12r^2} \right)(\sin^2\theta+ n^2 \delta^2) > 1-\frac{1}{n^2}\,.
\ee
At the tips, this condition implies that we cannot set the regulator $\delta $ to zero for $n>1$.

Now the von Neumann entropy is obtained from the modified R\'enyi entropy $\t S_n$ in the limit $n \to 1$; see (\ref{eq:St-def}), (\ref{eq:St-S}). We will do this by evaluating the stress tensor, and relating its integral to the $r$-derivative of the entropy, as discussed around (\ref{eq:ZnTrTSn}):
\be
r S'(r) =- \lim_{n \to 1}(1-n \partial_n)\,\int d^2x\,\sqrt{g}\,\langle\text{tr}\,T \rangle \,.
\ee
For simplicity, we will take $n\to 1$ from below to enable us to send $\delta\to 0$.
Our solution then takes the form (for $\eta=1$, which we consider for the remainder of this section)
\bea\label{eq:Tvevs2}
\langle T^\theta_\theta \rangle&=& \frac{1}{\pi \lambda} \left(1- \sqrt{1 + \frac{\lambda c}{12 r^2} \left( 1+ \epsilon \frac{C_{dS}}{\sin^2\theta}\right) } \right) \\
\langle T^\phi_\phi \rangle&=&\frac{1}{\pi \lambda} \left(1- \frac{1+ \frac{\lambda c}{12 r^2}}{\sqrt{1 + \frac{\lambda c}{12 r^2} \left( 1+ \epsilon \frac{C_{dS}}{\sin^2\theta}\right) }}   \right) \nonumber\,.
\eea
As mentioned above, there are two possible branches, and we took the one that gives the right CFT limit as $\lambda  \to 0$.
Here $C_{\text{dS}}$ can be understood as the integration constant for the bulk differential (conservation) equation, and we have introduced the parameter
\be\label{eq:ep}
\epsilon \equiv 1-n \ll 1\,,
\ee
which at the end we want to send to 0 for the calculation of the EE. 
The constant $C_{\text{dS}}$ is fixed by matching to the contribution from the conical deficits derived above, giving 
\be\label{eq:CdS}
C_{\text{dS}}=2\,.
\ee

We are now ready to evaluate the entropy. From (\ref{eq:Tvevs}) or (\ref{eq:Tvevs2}),
\be\label{eq:trTdS}
n \partial_n \langle \tr T \rangle  = \frac{c}{3}\,\frac{\lambda c}{48\pi}\,\frac{1-n^2}{n^4}\frac{1}{r^4 \sin^4 \theta}\,\frac{1}{\left(1+ \frac{\lambda c}{12 r^2}\left(1+ \frac{2 \epsilon}{\sin^2 \theta} \right)\right)^{3/2}}\,.
\ee
This expression is valid at finite $n$. This vanishes at $\epsilon=0$ or $n=1$, except at the endpoints $\theta=0, \pi$. So the contributions to the entropy integral
\be
r S'(r) = r^2 \,\lim_{n \to 1}\,(2 \pi n)\,\int_0^\pi d\theta\,\sin \theta\,n \partial_n \langle \tr T(\theta) \rangle
\ee
localize at the endpoints. Taking into account both endpoints,
\bea\label{eq:SdS}
r S'(r) &=& 2 \times 2\pi r^2\,\int_0^{\theta_0 \ll 1}\,d\theta \sin\theta\, \partial_n \langle \tr T(\theta) \rangle|_{n \to 1} \nonumber\\
&=& \frac{c}{3}\, \frac{1}{\sqrt{1+ \frac{\lambda c}{12 r^2}\left(1+ \frac{2 \epsilon}{\theta_0^2} \right)}}\Big|_{\epsilon=0}\nonumber\\
&=&  \frac{c}{3}\, \frac{1}{\sqrt{1+ \frac{\lambda c}{12 r^2}}}\,.
\eea
We note the order of limits here:  in passing to the last line we took $n\to 1$ at fixed $\theta_0$. Importantly, the shift in the denominator that we have derived here regulates the endpoints of the integral, leading to a cancellation of the $\epsilon$ factor in the numerator. The shift is a nonperturbative effect, derived from the trace flow and conservation equations.  Eq.~(\ref{eq:SdS}) reproduces the answer found in~\cite{Donnelly:2018bef}, showing explicitly how the nonvanishing contributions to the entropy localize at the endpoints.

Another interesting aspect of this result is that we recover the CFT answer by taking $\lambda \to 0$ in (\ref{eq:SdS}), $r S'(r) = c/3$. In our approach, this does not come from the delta-function singularities at the endpoints --we could solve the trace flow equation without them away from the endpoints, and instead they fixed the integration constant $C_{\text{dS}}$ in (\ref{eq:CdS}). Physically, $T \bar T$ is smoothing out the conical deficits, as we just saw explicitly in the endpoint calculation (\ref{eq:SdS}).
Below in \S\ref{sec:bSSA} we will argue that these properties also hold for an interval in Poincar\'e space. The localization property of the $T \bar T $ effects will then allow us to conformally map the dS and Poincar\'e cases, leading to an all-orders result for the interval entanglement entropy which agrees on both sides of the duality, analyzed independently.

\section{Cut off AdS/Poincar\'e case:  violation of boosted strong subadditivity}\label{sec:bSSA}

In this section we study the case of a $T \bar T$ deformed CFT in flat space, which is dual to cutoff AdS with Poincar\'e slices. In order to probe the effects of $T \bar T$, we will study the entanglement entropy. In particular, we will focus on the strong subadditive inequality (SSA), which is very sensitive to the locality properties of the algebras of regions. Our main result will be that $T \bar T$ violates the boosted version of SSA, something that we will exhibit independently in the field theory and gravity descriptions. We will trace this to the fact that in the deformed theory the algebra should not be associated to causal domains, but rather to smaller regions that we will characterize. The precise agreement between the gravity and field theory calculations is one of our main results, providing a nontrivial test of the duality in the non-symmetric cases that we will consider.

Before proceeding to the calculations and implications of this, let us review some basic aspects of algebras and the SSA in QFT, as well as the setup that we will use.

\subsection{Algebras and strong subadditivity in QFT}\label{subsec:SSAdef}

Given a spacelike region $R$, the domain of dependence $D[R]$ is the set of all points $p$ in spacetime such that every inextendible causal curve through $p$ intersects $R$. For intervals in 2d Minkowski spacetime, these are causal diamonds. Given causal evolution,\footnote{As discussed further in \S\ref{sec:subregion}, the AdS/Poincar\'e and (A)dS/dS cases have geodesic propagation fastest along the boundary rather than through the bulk, indicating that causality holds up to order $c^0$.} we expect that an operator in the domain of dependence belongs to the algebra of the region. So it is natural to associate an algebra of operators to $D[R]$, and we will denote it by $\mc A_{D[R]}$~\cite{Haag:1992hx}. Locality implies that if one spacetime region is included in another one, the corresponding algebras obey the same inclusion:
\be\label{eq:locality}
D[R] \subseteq D[\t R]\;, \;\;\mc A_{D[R]} \subseteq \mc A_{D[\t R]}\,.
\ee

In order to introduce the strong subadditive inequality, let us review the construction of the relative entropy. Consider a spacetime region $U$ with its corresponding algebra $\mc A_U$, and two states or density matrices $\rho_U, \sigma_U$. The relative entropy between the states is
\be\label{eq:Srel-def}
S(\rho_U | \sigma_U)= \tr \left(\rho_U\, \log(\rho_U \sigma_U^{-1})\right)\,.
\ee
This is a measure of the distinguishability between the states. The crucial property of relative entropy is its monotonicity: for two regions
\be\label{eq:locality2}
U \subseteq \tilde U\;,\; \mc A_U \subseteq \mc A_{\t U}
\ee
we have
\be\label{eq:monotonicityrel}
S(\rho_U | \sigma_U) \leq S(\rho_{ \tilde U} | \sigma_{\tilde U})\,.
\ee
See~\cite{Witten:2018lha} for a recent review with references to the original works. Intuitively, on a bigger algebra we can perform more measurements in order to distinguish the two states, and so the relative entropy should increase with the region. 

We are now ready to state the SSA. We take three regions $X, Y, Z$, a state $\rho_{XYZ}$ defined on the union, and its restrictions to the smaller subsystems. Monotonicity of the relative entropy implies
\be
S(\rho_{XYZ}|\rho_X \otimes \rho_{YZ}) - S(\rho_{XY}| \rho_X \otimes \rho_Y) \ge 0\,,
\ee
because the second term is obtained by doing a partial trace over $Z$, namely $\mc A_{XY} \subseteq A_{XYZ}$. Rewriting this using (\ref{eq:Srel-def}) gives the SSA for the von Neumann entropies,
\be\label{eq:SSA0}
S(\rho_{XY}) + S(\rho_{YZ}) - S(\rho_{XYZ}) - S( \rho_Y) \ge 0\,.
\ee
So the SSA measures the decrease of distinguishability in a 3-partite system under a partial trace. We will often rewrite this as
\be\label{eq:SSA-def}
S(A) + S(B) - S(A \cup B)- S(A \cap B)\ge 0\,,
\ee
with $A = XY,\, B=YZ, \,A\cup B=XYZ, \,A \cap B= Y$.

When the inequality is saturated, the state in $A \cup B$ can be reconstructed from the knowledge of the state in the smaller subsystems $A, B$~\cite{hayden2004structure}. This is called a Markov state, in analogy to classical Markov chains. Markov states play an important role in the structure of quantum field theory: the vacuum of a CFT saturates (\ref{eq:SSA-def}) for regions whose boundaries lie on a light-cone~\cite{Casini:2017roe}. 

\begin{figure}[h!]
\begin{center}  
\includegraphics[width=1.\textwidth]{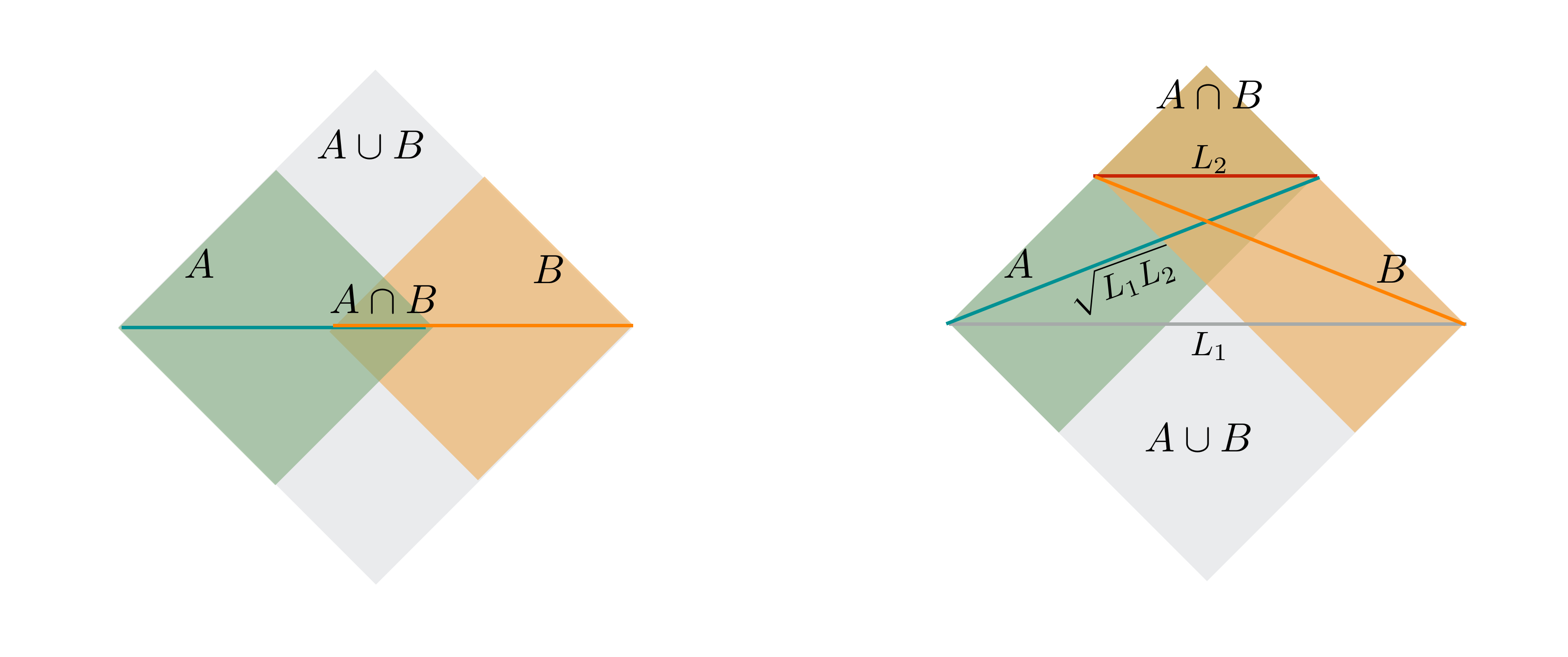}
\captionsetup{width=0.9\textwidth}
\caption{Setup for the spatial SSA (left) and boosted SSA (right). The causal domains of the intervals are also shown. In both cases, the union and intersection are spatial intervals with lengths $L_1$ and $L_2$ respectively.}
\label{fig:bSSA}
\end{center}  
\end{figure}

We will analyze two setups for the SSA, shown in Fig.~\ref{fig:bSSA}. The spatial SSA is illustrated in the left panel of the figure. Here all the intervals are spatial ($t=0$); their corresponding causal domains are also shown. We denote the length of $A \cup B$ by $L_1$, and that of $A \cap B$ by $L_2$; $A$ and $B$ both have length $(L_1+L_2)/2$. Then (\ref{eq:SSA-def}) reads
\be\label{eq:spatialSSA}
S(A) + S(B) - S(A \cup B)- S(A \cap B) = 2 \,S\left(\frac{L_1+L_2}{2}\right) - S(L_1)- S(L_2) \ge 0 \,.
\ee
For $L_1 = L + \delta L, L_2 = L-\delta L$ and $\delta L \to 0$, this becomes an infinitesimal inequality
\be
L^2\,S''(L) \le 0\,.
\ee

We will also study a stronger version of SSA, obtained by boosting some of the intervals~\cite{Casini:2012ei}; the setup is given in the right panel of Fig.~\ref{fig:bSSA}. $A$ and $B$ have their endpoints on a light-cone, such that $A \cup B$ and $A \cap B$ are purely spatial intervals, of length $L_1$ and $L_2$. In Minkowski geometry, $A$ and $B$ have length $\sqrt{L_1 L_2}$. The SSA now requires
\be\label{eq:bSSA-finite}
2\,S\left(\sqrt{L_1 L_2} \right)- S(L_1)- S(L_2) \ge 0\,,
\ee
and the infinitesimal version $L_2 \to L_1$ becomes
\be\label{eq:boostedSSA}
L^2 S''(L) +L S'(L) \le 0\,.
\ee
(This was used in~\cite{Casini:2012ei} to prove the $c$-theorem.) 

We will find that spatial SSA is respected by $T \bar T$, while the boosted inequality, in both finite and infinitesimal versions, is violated. As we just reviewed, the SSA is a consequence of monotonicity of the relative entropy, which follows from the locality property (\ref{eq:locality}). So for the deformed theory in the boosted setup, the locality property should fail.

\subsection{Gravity side}\label{subsec:bSSAgr}

Let us start with the simpler gravity side calculation of the entanglement entropy. We have the $AdS_3$ metric with Poincar\'e slices and a finite cutoff at $z=z_c$:
\be
ds^2 = \ell^2\frac{dz^2-dt^2+dx^2}{z^2}\;,\;z \ge z_c\,.
\ee
At the cutoff surface we consider an entanglement interval $x \in (-L/2, L/2)$.  We tentatively assume that the bulk path integral arguments \cite{Lewkowycz:2013nqa,Faulkner:2013ana}\ for holographic entanglement entropy \cite{Ryu:2006bv,Hubeny:2007xt} and its $1/c$ corrections persist in the cut off case, although we should not take for granted the factorization of the Hilbert space into spatial subregions (given a lattice regularization of the path integral) as in local quantum field theory.  In \S\ref{sec:subregion}\ we will delve into the general issues which arise in the deformed theory.  However, all of our results, including both-sides calculations of entanglement entropies, will support the conjecture that the Ryu-Takayanagi (RT) and Hubeny-Randamani-Takayanagi (HRT) surfaces' areas define a meaningful entropy under a division of the system that can be associated in a certain sense with a putative subregion $R$, with the corresponding algebra ${\cal A}_{R_s}$ differing in some ways from the pure (undeformed) QFT case.

The minimal Ryu-Takayanagi (RT) surface~\cite{Ryu:2006bv} is given by extremizing the area
\be
S=\frac{1}{4 G_N} 2\ell \int_{z_c}^{z_t} \frac{dz}{z} \sqrt{1+ (\partial_z x)^2}
\ee
with boundary conditions
\be
x(z_c)=L/2\;,\;\partial_z x(z_t) \to \infty\,.
\ee
Here $z_t$ is the turning point of the minimal curve.

The result is well-known; it is a half-circle
\be\label{eq:halfcircle}
x^2+z^2 = (L/2)^2+z_c^2\,.
\ee
The on-shell integral for the extremal area then reads
\be\label{eq:Slocalization}
S= \frac{1}{2 G_N}\,\int_{z_c}^{z_t}\,\frac{dz}{z}\,\frac{1}{\sqrt{1-z^2/z_t^2}}\,,
\ee
and it is easy to check that all the contribution to the entropy comes from $z=z_c$ or, equivalently in the $x$ variable, from the endpoints of the entangling region.
The entropy evaluates to
\be\label{eq:S2d}
 S(L)= \frac{c}{3} \log \left(\frac{L}{2z_c}+ \sqrt{1+ \left(\frac{L}{2z_c} \right)^2} \right)\,.
\ee
We used $c= \frac{3\ell}{2G_N}$.
As $z_c \to 0$, this recovers the CFT answer $S =\frac{c}{3} \log(L/z_c)$, but here we are interested in keeping $z_c$ finite. The entropy is then finite and, in particular, $S(0)=0$.

Let us study the SSA with the goal of quantifying the effects of $T \bar T$. Plugging (\ref{eq:S2d}) into the spatial SSA (\ref{eq:spatialSSA}), we find that this inequality is respected. In this sense, $T \bar T$ is not having an important effect on local algebras that differ by spatial contractions. In contrast, we find that the boosted SSA, both in its finite (\ref{eq:bSSA-finite}) and infinitesimal (\ref{eq:boostedSSA}) forms, is violated.\footnote{By Lorentz invariance, the entropy can only depend on the invariant length of the interval. So we can still use (\ref{eq:S2d}) for a boosted interval.} In particular, the infinitesimal SSA combinations become
\bea\label{eq:boostedSSA1}
\text{spatial SSA:}\;\;L^2 S''(L)&=&-\frac{c}{3}\,\frac{1}{\left(1+\frac{z_c^2}{(L/2)^2}\right)^{3/2}} \le 0 \nonumber\\
\text{boosted SSA:}\;\;L^2 S''(L) +L S'(L) &=&\frac{c}{3} \,\frac{z_c^2}{(L/2)^2}\, \frac{1}{\left(1+\frac{z_c^2}{(L/2)^2}\right)^{3/2}} \ge 0\,.
\eea
Note that a 2d CFT just saturates the boosted SSA  (\ref{eq:bSSA-finite}), so this combination is dominated by the deformation. What we are finding is that $T \bar T$ leads to a violation of boosted SSA. In line with this, the running C-function
\be
C(L) = L S'(L)= \frac{c}{3}\,\frac{1}{\sqrt{1+4z_c^2/L^2}}\,,
\ee
exhibits a volume-law $C(L) \approx \frac{c}{6} \frac{L}{z_c}$ for small $L$. Volume-law entanglement in the vacuum is a signature of nonlocal interactions.

These properties of the $T \bar T$ deformed theory are quite striking. On the one hand, spatial SSA (\ref{eq:bSSA-finite}) is always satisfied for any sizes $L_1$ and $L_2$. This suggests that at $t=0$ we have a local quantum mechanical system, and that we can specify the subsystem associated to a bounded region $R$ in the usual way in terms of the algebra of operators in the region. On the other hand, the failure of the boosted SSA implies that locality is violated under a relative boost between regions.

\subsubsection{HRT surface for boosted intervals}

In order to develop intuition for this apparent violation of the inclusion property (\ref{eq:locality}), let us look in more detail into the extremal surfaces associated to the intervals in the boosted SSA construction. For concreteness, we focus on the interval $A$ in the right panel of Fig.~\ref{fig:bSSA}, with endpoints
\be\label{eq:intA}
(t, x) = \left(0, - \frac{L_1}{2}\right), \,\left(\frac{L_1-L_2}{2},\frac{L_2}{2}\right)\,.
\ee
We look for a geodesic that connects these points at $z=z_c$, obtained by extremizing
\be\label{eq:Ageodesic}
S=\frac{\ell}{4 G_N}\, 2 \int_{z_c}^{z_t}\,\frac{dz}{z} \sqrt{1+ x'(z)^2-t'(z)^2}\,.
\ee

The momenta of $x(z), t(z)$ in (\ref{eq:Ageodesic}) are conserved, which allows for a straightforward integration the equations of motion.
The integration constants are fixed by the initial conditions (\ref{eq:intA}), obtaining
\bea\label{eq:HRT1}
t(z) &=& \frac{L_1-L_2}{4} \left( 1 \pm \sqrt{1-4 \frac{z^2-z_c^2}{L_1 L_2}} \right) \nonumber\\
x(z) &=& - \frac{L_1}{2} + \frac{L_1+L_2}{4} \left( 1 \pm \sqrt{1-4 \frac{z^2-z_c^2}{L_1 L_2}} \right) \,.
\eea
The two branches connect at  the turning point
\be
z_t^2 = z_c^2 + \frac{L_1 L_2}{4}\,.
\ee

We can now compare the extremal surfaces of the spatial interval $A \cup B$, and the boosted one $A$, and their entanglement wedges. \S\ref{sec:subregion} will be devoted to a detailed analysis of the entanglement wedge (EW) and causal wedge (CW), and their connection to local algebras. For our purpose here, we will only discuss the following aspect.
In the spatial case, $L_1=L_2$, and (\ref{eq:HRT1}) gives the RT surface
\be\label{eq:AcupB}
t=0\;,\;x^2+z^2 = \left(\frac{L}{2} \right)^2+z_c^2\,.
\ee
The EW is the domain of dependence of the region bounded by (\ref{eq:AcupB}) and the cutoff surface $z=z_c$. It is delimited by the cone
\be
|t| = \sqrt{ \left(\frac{L}{2} \right)^2+z_c^2}- \sqrt{x^2+z^2}\,,
\ee
obtained by shooting light-rays from the RT surface to the boundary. Note that at the boundary $z=z_c$, this delimits a region
\be\label{eq:boundaryEW}
|t| = \sqrt{ \left(\frac{L}{2} \right)^2+z_c^2}- \sqrt{x^2+z_c^2}\,,
\ee
that is inside the causal diamond.
 Now, it is clear that this EW does not contain the HRT surface of the boosted interval (\ref{eq:HRT1}). As a consequence, the EW of the boosted interval $A$ is not contained in the EW of $A \cup B$.

In theories with gravity duals, it is natural to associate the algebra of operators to the EW. For a boundary region $R$, we will denote this algebra by $\mc A_{R_s}$.
In this case the EW is contained within the CW (see \S\ref{sec:subregion} for more details), so the spacetime region (\ref{eq:boundaryEW}) defined at the boundary will be strictly smaller than the causal diamond.  The holographic calculation that we just performed then shows that for the setup just considered, 
\be\label{eq:boostedA}
\mc A_{A_s} \,\not \subseteq\, \mc A_{(A \cup B)_s}\, ~~~~ {\text{when}} \not\exists ~ \Sigma|_{A, A\cup B \in \Sigma}  
\ee
with $\Sigma$ a spacelike surface.  
This can account for the failure of SSA. Even though $D[A] \subseteq D[A \cup B]$, the corresponding algebras do not obey the inclusion property. In other words, new operators are being accessed in the deformed theory when there is a relative boost between the regions, as in (\ref{eq:boostedA}). When there is no relative boost between the regions, we have  $\mc A_{A_s} \, \subseteq\, \mc A_{(A \cup B)_s}$, and correspondingly the spatial SSA is respected.

This failure of inclusion under a relative boost is reminiscent of the type of target spacetime non-locality that occurs in string theory \cite{Susskind:1993aa, Dodelson:2017emn, Dodelson:2017hyu}.  
In that context, under a relative boost, an observer develops sharp enough time resolution to detect a large variance of the high frequency modes of the string embedding coordinates.  This is somewhat analogous to the increase in accessible operators in the algebras just described. That said, we should stress that this is not a direct analogy:  in particular, the sign of the $T\bar T+\dots$ deformation we are working with here is the opposite of the sign for which the dressed version of a set of free scalars gives the standard Nambu-Goto string action \cite{Dubovsky:2012wk}.

\subsection{Deformed CFT side}\label{subsec:defCFT}

We will now analyze the entanglement entropy using the deformed CFT. 
Since~\cite{McGough:2016lol, Kraus:2018xrn, Gorbenko:2018oov}
\be
\frac{z_c^2}{L^2} = \frac{\lambda c}{12L^2}\,,
\ee
we can rewrite (\ref{eq:S2d}) in field theory terms as
\be\label{eq:EEinterval}
L S'(L)= \frac{c}{3}\,\frac{1}{\sqrt{1+ \frac{\lambda c}{3L^2}}}\,.
\ee
Our goal is to reproduce this expression with field theory methods. To this end, we note the striking similarity between (\ref{eq:EEinterval}) and the $dS_2$ result (\ref{eq:SdS}) -- both agree if we identify $L=2r$.\footnote{The similar holographic behavior of the interval and dS entropies was also noticed by~\cite{Park:2018snf}.} By generalizing the CHM map~\cite{Casini:2011kv} to the $T \bar T$ deformation, we will show that this coincidence is not accidental.

We consider a spatial interval of length $L$, and compute the entropy by the replica method. As reviewed in \S\ref{subsec:DS}, $L S'(L)$ is related to the integral of the stress tensor trace in the replicated space, (\ref{eq:ZnTrTSn}). So we begin by analyzing the expectation value of the stress tensor.

In the limit $n \to 1$ we recover flat space, and in vacuum
\be
\langle T \rangle_{\lambda,n=1} =\langle \bar T \rangle_{\lambda,n=1} =0\,.
\ee
As we review below, these expectation values are non-vanishing at order $\epsilon=1-n$. So the trace flow equation (\ref{Teqs}) with $\eta=1$ gives
\be
\langle \tr\, T \rangle = - \frac{\pi \lambda}{2} \left(\langle T^{ab} \rangle \langle T_{ab} \rangle - (\langle T^a_a \rangle)^2\right) \sim \epsilon^2
\ee
when $\epsilon \to 0$, and away from the conical singularities.\footnote{We will include their effect shortly.}
This implies that the contribution to the EE has to come from the endpoints $\rho \to 0$ of the entangling region {(with $\rho$ the radial direction in polar coordinates near the endpoint)}: the $n \rightarrow 1$ and $\rho \rightarrow 0$ limits need not commute.
If these two limits commuted, then $\partial_L S(L)$ would be independent of $\lambda$, since there would be no contributions from $T \bar{T}$ at finite $\lambda$. Holographically, this localization is seen in the on-shell integral (\ref{eq:Slocalization}) for the entropy, which receives its contribution only from $z=z_c$.
 
To obtain $ \langle \tr\, T \rangle$ near the endpoints, we note that as $\rho \to 0$ we have approximate rotational invariance in the vacuum state. This symmetry implies that $\langle T_{\rho \phi} \rangle =0$ near the endpoints. Solving the trace flow and conservation equations with this ansatz (see the next subsection for more details) gives
\be\label{eq:trTansatz}
\langle \tr\, T \rangle= \frac{2}{\pi \lambda} \left(1- \frac{\rho^2+ \t C}{\rho \sqrt{\rho^2+2 \t C}} \right)\,.
\ee
We have not included the curvature of the conical singularities. As in \S\ref{subsec:DS}, the integration constant $\tilde C$ is determined by 
including the (smoothed) curvature and requiring regularity as $\rho \to 0$. 

In order to make manifest the limits $\langle \tr\, T \rangle \sim O(\lambda)$ as $\lambda \to 0$ and $\langle \tr\, T \rangle \sim O(\epsilon^2)$ as $n \to 1$, let us redefine $\tilde{C}=\epsn\, \frac{\lambda c}{12} \,C(\frac{\lambda}{L^2})$. Then
\be\label{eq:TCform}
n \partial_n \langle \tr\, T \rangle |_{n \to 1}=\epsilon\, \frac{c}{3}\,\frac{\lambda c}{24\pi}\,\frac{C(\frac{\lambda}{L^2})^2}{ \rho\left(\rho^2+\epsilon \frac{\lambda c}{6} C(\frac{\lambda}{L^2}) \right)^{3/2}}\,.
\ee
Taking into account the contributions from both endpoints, the result for the entropy is
\bea\label{eq:Sint}
L S'(L) &=& 2 \times \lim_{n\to 1} (2\pi n) \int_0^{\rho_0 \ll L} \rho d\rho\, n \partial_n \langle \tr\,T \rangle\nonumber\\
&=& \frac{c}{3}\,C(\frac{\lambda}{L^2})\, \frac{1}{\sqrt{1+\epsilon \frac{\lambda c}{6 \rho_0^2} C(\frac{\lambda}{L^2})}}\Big|_{\epsilon\to 0} \nonumber\\
&=& \frac{c}{3}\,C(\frac{\lambda}{L^2})\,.  
\eea  
Here, $\rho_0\ll L$ is a small value consistent with the near-endpoint estimates we have made for $\tr(T)$.  We note an order of limits issue here:  in passing to the last line we took $n\to 1$ at fixed $\rho_0$, rather than taking first a near-endpoint limit $\rho_0^2/\lambda \to 0$.  We encountered a similar situation above in (\ref{eq:SdS}). Note that the shift in the denominator that we have derived here regulates the $\rho=0$ endpoint of the integral, leading to a cancellation of the $\epsilon$ factor in the numerator. We will calculate $C(\lambda)$ exactly below in \S \ref{subsec:allorders}.

From the solution of the trace flow equation we can also understand the behavior of the twist operators in the deformed theory. Changing from spherical to complex coordinates $(z, \bar z)$, the expectation value of $T(z)$ near the endpoint, which we choose at $z= \bar z =0$, obtains
\be\label{eq:Tz}
\langle T(z) \rangle = - \epsilon \frac{c}{24\pi z^2}\,\frac{C(\lambda)}{\sqrt{1+ \epsilon \frac{\lambda c}{6|z|^2}C(\lambda)} }\,,
\ee
for $n\sim 1$. We will prove shortly that the undeformed result as $\lambda \to 0$ is $C(0)=1$, in which case this reproduces the known CFT answer near an endpoint of the entangling region~\cite{Calabrese:2009qy}. However, in the deformed theory we again find a nonperturbative denominator shift that changes the behavior as $z \to 0$. This can be translated into the OPE with the twist operators $\mc T_n$, recalling that for an interval $(u, v)$, we have~\cite{Calabrese:2009qy}
\be\label{eq:OPE}
\langle T(z) \rangle_{\mc R_n}= \frac{\langle \mc T_n(u,0) \tilde {\mc T}_n(v,0) T(z)\rangle_{\mathbb C}}{\langle \mc T_n(u,0) \tilde{ \mc T}_n(v,0) \rangle_{\mathbb C}}\,,
\ee
with the left hand side evaluated on the replicated space, and the right hand side calculated in flat space. From (\ref{eq:Tz}) and (\ref{eq:OPE}), we see that the $T \bar{T}$ deformation changes the leading divergence in the OPE between the stress tensor and the twist operator for the conical singularity. This is a new instance in which the non-local behavior of the deformation is exhibited.

\subsubsection{Leading calculation}\label{subsec:leading}

Let us explain  how the leading result $C(\frac{\lambda}{L^2})=1$ comes from the trace flow equation. 

The effects of the curvature singularities in the replicated geometry can be taken into account as in \S\ref{subsec:DS}; near one of the endpoints,
\be
ds^2 = \frac{\rho^2 + n^2 \delta^2}{\rho^2+\delta^2}\,d\rho^2 + n^2 \rho^2 d\phi^2\;,\;\delta \to 0\,,
\ee
and the Ricci scalar evaluates to
\be
\mc R^{(2)} = -2 (n^2-1) \frac{\delta^2}{(\rho^2+n^2 \delta^2)^2}\,.
\ee
We will work with finite but small $\delta$, solve the conservation and trace flow equations, impose regularity at $\rho=0$, and then take $\delta \to 0$. The limit will turn out to be finite in the presence of the $T \bar T$ deformation.

The trace flow equation in polar coordinates reads
\be
T^\rho_\rho + T^\phi_\phi = - \frac{c}{24\pi} {\mc R}^{(2)} - \frac{\eta-1}{\pi \lambda} + \pi \lambda \left( T^\rho_\rho T^\phi_\phi-n^2 \rho^2(T_\rho^\phi)^2\right)\,.
\ee
The smoothing parameter $\delta$ only matters in the scalar curvature above, while we neglect it the other terms and in the conservation equations. By rotational invariance near the endpoint, $T_\rho^\phi=0$ in the vacuum. The remaining conservation equation simplifies to
\be\label{eq:consT3}
\partial_\rho T^\rho_\rho  +\partial_\phi T^\phi_\rho+\frac{1}{\rho} (T^\rho_\rho - T^\phi_\phi) =0\,.
\ee

Following steps similar to those in \S\ref{subsec:DS} gives (\ref{eq:TCform}) with $C(\lambda/L^2)=1$. This matches with the one interval entropy $L S'(L)=c/3$ of the undeformed CFT. It is interesting that in order to reproduce the $\lambda=0$ answer, one needs to keep $\epsn \lambda$ finite until the end of the calculation; this illustrates rather explicitly that finite $\lambda$ is acting like a regulator for the conical singularity. 

The details are interesting to display directly in the Poincar\'e case.  
The solution that is regular as $\rho \to 0$ for finite $\delta$ is
\be
\langle T^\rho_\rho \rangle = \frac{1}{\pi \lambda} (1+U)\;,\; \langle T^\phi_\phi \rangle =\frac{1}{\pi \lambda} (1+V)
\ee
with
\be
U = - \frac{\sqrt{\eta(\rho^2 + n^2 \delta^2)+ \frac{\lambda c}{12}(\frac{1}{n^2}-1)}}{\sqrt{\rho^2 + n^2 \delta^2}}\;,\;V=\frac{\frac{\lambda c}{24} {\mc R}^{(2)}(\rho) + \eta}{U}\,.
\ee
Finally, taking $\delta \to 0$, the resulting stress tensor is
\bea\label{eq:leadingT}
\langle T^\rho_\rho\rangle &=& \frac{1}{\pi \lambda}\left(1-  \sqrt{\eta+\frac{\lambda c}{12 \rho^2}(\frac{1}{n^2}-1)} \right) \nonumber\\
\langle T^\phi_\phi \rangle &=& \frac{1}{\pi \lambda}\left(1-  \frac{1}{\sqrt{\eta+\frac{\lambda c}{12 \rho^2}(\frac{1}{n^2}-1)}} \right)\,.
\eea
From this, setting $\eta=1, n = 1-\epsilon$ and taking $\epsilon \to 0$, we read off the trace
\be\label{eq:aitorA}
n \partial_n \langle {\text tr}\,T \rangle |_{n \to 1} =\epsilon \frac{\lambda c^2}{72 \pi}\,\frac{1}{\rho \left( \rho^2 +\epsilon \frac{\lambda c}{6}\right)^{3/2}}\,.
\ee
This is of the form (\ref{eq:TCform}) with $C(\lambda/L^2)=1$.
Again, this indicates that $\lambda$ itself regulates the singularities, and it is consistent to maintain $\delta\ll \sqrt{\lambda}$ throughout the calculation, so that the regulator does not interfere with the effects of the deformation parameter $\lambda$.

\subsubsection{All-orders result for the entropy}\label{subsec:allorders}

Let us now consider the higher-order contributions in $\lambda$. 
What makes the $T \bar T$ deformation special from the point of view of the entanglement entropy is that all the $\lambda$ dependence comes from the endpoints, where there is an enhanced rotational symmetry. This behavior is expected to be universal, but the challenge is to extract the dependence on $\lambda/L^2$, that obviously cares about the total size of the region.

Previously we remarked on the similarity between the de Sitter result (\ref{eq:SdS}) and the expression (\ref{eq:EEinterval}) that we want to prove. This motivates us to consider the conformal transformation that maps the domain of the dependence of the interval in Poincar\'e space to the de Sitter static patch. This was used in~\cite{Casini:2011kv} for CFTs, establishing the connection between the trace anomalies on the sphere and the universal logarithmic term in the entropy, as well as providing a derivation of the Ryu-Takayanagi formula for spherical regions. In general, this conformal map is not very useful for QFTs away from fixed points:
non-marginal couplings become spacetime dependent under the conformal transformation. However, as we will shortly review, the conformal transformation goes to the identity at the endpoints. The special localization property of $T \bar T$ near the endpoints would then imply that the $\lambda$ dependence is not affected by such a transformation. We will now analyze how this comes about.

We recall that the conformal map from the domain of dependence of the interval $x^1 \in (-r, r)$ to the static patch is~\cite{Casini:2011kv}
\be
x^0 = r \frac{\sin \theta\, \sinh (\tau/r)}{1+ \sin \theta\, \cosh(\tau/r)}\;,\;x^1 = r \frac{\cos \theta}{1+ \sin \theta\, \cosh(\tau/r)}\,.
\ee
This transforms the metric as
\be
ds^2 =-(dx^0)^2 + (dx^1)^2 = \Omega(\theta, \tau)^2\, (-\sin^2\theta\, d\tau^2+r^2\,d\theta^2)
\ee
with conformal factor
\be
\Omega(\theta, \tau)=\frac{1}{1+ \sin \theta\, \cosh(\tau/r)}\,.
\ee
The size of the interval 
\be\label{eq:LR}
L=2r
\ee 
becomes twice the radius of de Sitter. The domain of dependence of the interval maps to the static patch:
\bea
\tau \to \pm \infty\;:\; (x^0, x^1) \to (\pm r, 0)\;;\;\theta \to 0, \pi\;:\; (x^0, x^1) \to (0, \pm r)\,.
\eea
We will also need the euclidean version; after analytic continuation from $\tau/r$ to $\phi$, the conformal factor becomes
\be\label{eq:Omegaeucl}
\Omega(\theta, \phi)= \frac{1}{1+\sin \theta\, \cos \phi}\,.
\ee
As anticipated, this becomes trivial at both endpoints: $\Omega(0, \phi) = \Omega(\pi, \phi)=1$.

We would like to argue for the equality of the interval and dS entropies from the deformed CFT side using this transformation. We start from the expression (\ref{eq:Sint}),
\be
L S'(L) = 2 \times \lim_{n\to 1} (2\pi n) \int_0^{\rho_0 \ll L} \rho d\rho\, n \partial_n \langle \tr\,T \rangle\,,
\ee
localized at the endpoint. We now apply the ($n$-sheeted version of the) conformal transformation (\ref{eq:Omegaeucl}) to de Sitter:
\be\label{eq:EEdS}
L S'(L) =  2 \times \lim_{n\to 1}  r^2 \int_0^{\theta_0 \ll 1} \theta d\theta\,d\phi\, n \partial_n \langle \tr\,T \rangle_{\tilde \lambda= \Omega_n^{-2}\lambda}\,,
\ee
with $r=L/2$ as found in (\ref{eq:LR}).
The conformal transformation has two effects: it produces the Weyl anomaly in de Sitter, and it makes the $T \bar T$ coupling $\tilde \lambda$ spacetime-dependent. 
So we end up with the new trace-flow equation
\be\label{eq:newtrT}
\tr\, T= -\frac{c}{24\pi}\,\frac{2}{r^2}+ \pi \lambda \,\Omega_n(\theta, \phi)^{-2} \left(T^\theta_\theta T^\phi_\phi-\sin^2 \theta\,  (T_\theta^\phi)^2\right)\,.
\ee
We have ignored the delta function curvature terms at the tips which, as before, will just fix the integration constant.

We have to solve the trace flow and conservation equations near the endpoint $\theta \to 0$ (there is a similar contribution from $\theta \to \pi$). The conformal factor always goes to the identity at the endpoints, $\Omega_n(0, \phi)=1$, even in this replicated case; the problem then reduces to the symmetric half-space division of \S\ref{subsec:DS}. Note that all the dependence on the size of the region $L=2r$ has been packaged into the Weyl anomaly, while the problem retains the rotational symmetry associated to the fact that only the near-endpoint region contributes in the original problem. The solution for $\partial_n \langle \tr T \rangle$ is then (\ref{eq:trTdS}), and following the same steps as around that equation leads to 
\be\label{eq:EEinterval2}
L S'(L)= \frac{c}{3}\,\frac{1}{\sqrt{1+ \frac{\lambda c}{3L^2}}}\,.
\ee
This reproduces precisely the holographic answer.

This result is valid at large $c$ and to all orders in (fixed) $\lambda c$.\footnote{Large $c$ was used in the de Sitter calculation for the factorization of $T \bar T$.} Obtaining an exact result for the entropy in a non-conformal theory is quite surprising and interesting in itself. As we just discussed, the reasons for this stem from the properties of the trace flow equation and the localization of the $T \bar T$ contributions at the endpoints of the entangling region. This exact all-orders match provides a very nontrivial test of the duality.
Finally, we note that on the deformed field theory side, we have computed $L S'(L)$ but not $S(L)$ itself.  
It will be interesting to explore further the integration constant in the field theory side, and its potential role as a parameter in the $\lambda$ deformed partition function.

Having reproduced the holographic result directly on the deformed CFT side, let us finally return to the discussion of SSA. As in \S\ref{subsec:bSSAgr}, we have that the spatial SSA is respected, while the boosted SSA is violated, to all orders in $\lambda c$ at large $c$. The differential version of the spatial SSA inequality is probing very short scales near the endpoints of the region. This suggests that we have spatial locality and a standard division of the system in terms of algebras of regions. However, the boosted SSA behavior is consistent with time evolution, and modular evolution for the complement, leading to highly nonlocal operators that cannot be reproduced from the local algebra of the region. We will discuss these effects in more detail in \S\ref{sec:subregion}. 

\section{Subregion dualities}\label{sec:subregion}

Originally, when considering holographic dualities for a boundary subregion, the entanglement wedge and causal wedge were proposed \cite{Headrick:2014cta}. We will
be interested in their fate in the deformed theories.  In this section we will review their properties in asymptotic AdS/CFT, and explain some essential new features that arise in the deformed theories.      

The entanglement wedge (EW)\footnote{Note that the entanglement wedge is usually defined as the domain of the dependence of the region between $D[R]$ and the RT surface, but in our situation, part of $D[R]$ might be time-like related with the RT surface, so we choose this more restricted definition of the entanglement wedge.} is defined as the domain of dependence of the region between the RT surface and the boundary region $R$.
This notion continues to exist in all the cutoff theories. 
The causal wedge (CW) is defined by the bulk points which can be reached by intersecting future directed and past directed rays thrown from the domain of dependence of the boundary region, $D[R]$.  This notion is tricky in the cutoff theories, as it is not always the case that signals propagate fastest along the boundary and hence $D[R]$ is not of clear significance \cite{Nomura:2018kji}.  However, in the cutoff (A)dS/dS as well as the AdS/Poincar\'e  theories, the de Sitter boundary preserves this feature of AdS/CFT, and we will ultimately focus on these cases for our study of causal wedge reconstruction. Again, in all versions of the deformed theory we can naturally study the EW case.   

\subsection{Algebras, subregions and causality}\label{subsec:asc}

We are working with a seed holographic CFT with emergent bulk locality, so that general relativity and bulk effective field theory apply down to the string scale.  
To leading order in $G_N$, bulk effective fields are free and one can use their wave equation and the extrapolate dictionary to write them in terms of boundary operators (in the boundary at large $c$ these operators are generalized free fields, GFF, which are gaussian fields). This allows us to map the algebra of EFT bulk operators in some given bulk Cauchy slice to the algebra of low energy boundary operators: ${\cal A}^{\text{bulk,EFT}}={\cal A}^{\text{bdy,GFF}}$.  

\subsubsection{On the definition of operators at a finite cutoff}  

In the deformed theories with a Dirichlet wall, we can characterize the boundary operators in terms of the radial field momentum as in \cite{Hartman:2018tkw}. This is a natural generalization of the dictionary for the boundary stress tensor in terms of the extrinsic curvature of the boundary \cite{McGough:2016lol}.  
 
In the presence of a cutoff surface, we can write a bulk operator in terms of cutoff boundary operators by using the wave equation. Because of causality, we write the bulk operator as a linear combination of boundary operators (one can think of this as a Bogoliubov transformation):
\begin{equation}\label{HKLLgen}
\Phi(X)=\int dx f(X|x) O(x)\,,
\end{equation}
where $X$ denotes the bulk point and $x$ the boundary point.  
We can obtain an expression for the kernel $f$ in terms of correlators using the fact that to leading order in $G_N$, the bulk fields are free, so we can just take the expectation value of the previous expression with $O(x')$ and formally invert the boundary two-point function\footnote{Here we are using the notation $\tilde{O}$ to denote the shadow operator. The shadow operator $\tilde{O}$ for an operator $O$, with dimension $\Delta$ and spin $J$ in CFT, is defined by a dual operator with dimension $d-\Delta$ and spin $J$, where $d$ is the spacetime dimension. See for instance \cite{SimmonsDuffin:2012uy,Karateev:2018oml}.}:
\begin{eqnarray}\label{eq:shadow}
f(X|x)&=&\langle \Phi(X) \tilde{O}(x) \rangle  \\ 
 \tilde{O}(x)&=&\int dx' O(x) (\langle O(x) O(x') \rangle)^{-1},\; \langle \tilde{O}(x) O(x')\rangle = \delta(x-x') \nonumber\,.
\end{eqnarray}

  There are two differences between our story and the HKLL situation~\cite{Hamilton:2005ju} where we have asymptotic $\text{AdS}$:
  here $f(X|x)$ generalizes the bulk-to-boundary correlator, containing a bulk-to-cutoff surface correlator; and the two-point function of boundary operators will depend $\lambda$. 
In these expressions, we have not yet constrained the behavior of $f(X|x)$ as $X$ goes to the boundary surface. In the presence of Dirichlet boundary conditions, we get the extrapolate dictionary by identifying the radial momenta at the cutoff boundary with $O(x)$: $\partial_N \Phi(X)|_\text{bdy}=O(x)$,
where $N$ is the normal direction to the boundary.
In general, the only complication of HKLL is doing the inverse of the boundary correlator, which is  only doable analytically if there is enough symmetry to diagonalize explicitly the wave equation. The kernel $f(X|x)$ should be though of as a distribution.  In Appendix~\ref{sec:HKLLappendix}\ we lay this out for a particular case of interest.  

\subsubsection{Operator algebras and inclusions}

Let us now discuss the algebraic properties from the point of view of the boundary theory.
The wedges CW and EW previously defined characterize bulk subregions to which we can associate algebras of operators in effective field theory. These algebras of operators are by now well understood. In the case of the causal wedge, since it is a boundary construction, it contains all low-energy local operators localized in $D[R]$: 
\be\label{eq:defACW}
{ \cal A}_\text{CW}={ \cal A}_{D[R]}=\lbrace {O(x) ,\; x \in D[R] \rbrace }\,.
\ee
The entanglement wedge is more complicated because it is a bulk construct, but as explained in \cite{Faulkner:2017vdd}, it is constructed from local operators in $R$ and their evolution with the density matrix: 
\be\label{eq:defAEW}
{\cal A}_\text{EW}={ \cal A}_{R_s}=\lbrace {\rho_R^{i s}\, O(x)\, \rho_R^{-i s} ,\; x \in R, \;s \in \mathbb{R} \rbrace }\,.
\ee

Here $\rho_R$ denotes the density matrix obtained by a division of the system which for the bulk effective field theory is a spatial division along the Ryu-Takayanagi extremal surface. 
In our deformed theories, this need not correspond to a local decomposition in the boundary theory.
At the same time, we should keep in mind that the Hamiltonian evolution is not generally local in $T\bar T$ deformed theories.  We will return to these aspects shortly, after deriving the commutator structure of our algebras.   

In the following, we will work directly with the boundary algebras ${\cal A}_{D[R]}, {\cal A}_{R_s}$ which make no reference to the bulk.  We would like to explain some properties that they satisfy in the case of interest with a large-radius bulk dual.   First we will review the standard setup, and then come to the deformations. 

 In asymptotic $\text{AdS}$, the causal wedge is claimed to be always contained in the entanglement wedge \cite{Headrick:2014cta, Wall:2012uf}. This implies that  ${\cal A}_\text{CW} \subset {\cal A}_\text{EW}$; in other words, in the effective theory there have to be non-local operators acting within $R$ that one is missing when considering $D[R]$. The natural generator of non-local operators in $R$ is the density matrix or modular hamiltonian $K_R=-\log \rho_R$ and, as shown in \cite{Faulkner:2017vdd}, the modular hamiltonian is enough to generate the operators in ${\cal A}_{R_s}$ which are non-local in $D[R]$.

\begin{figure}[h!]
\begin{center}  
\includegraphics[width=0.48\textwidth]{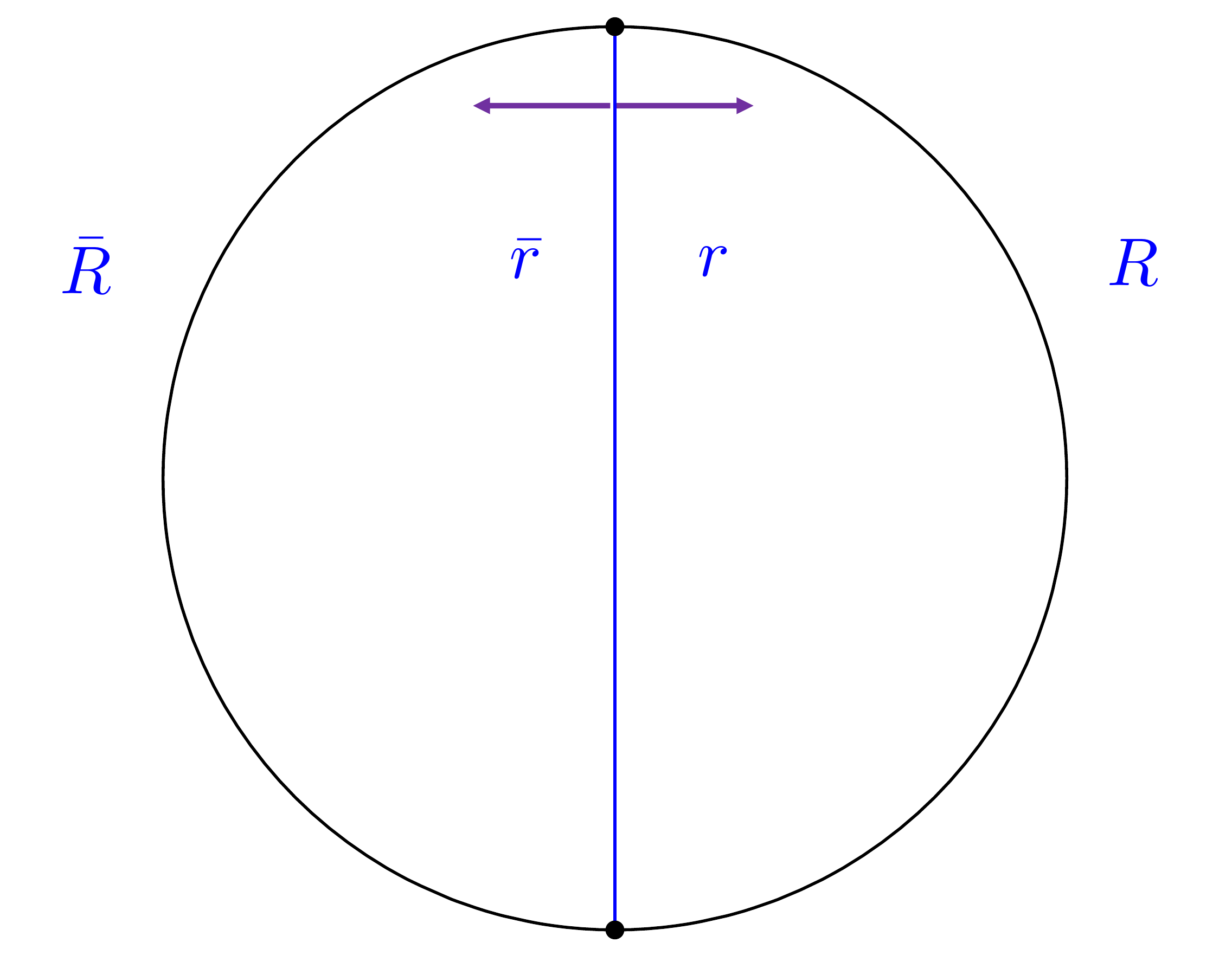}
\includegraphics[width=0.48\textwidth]{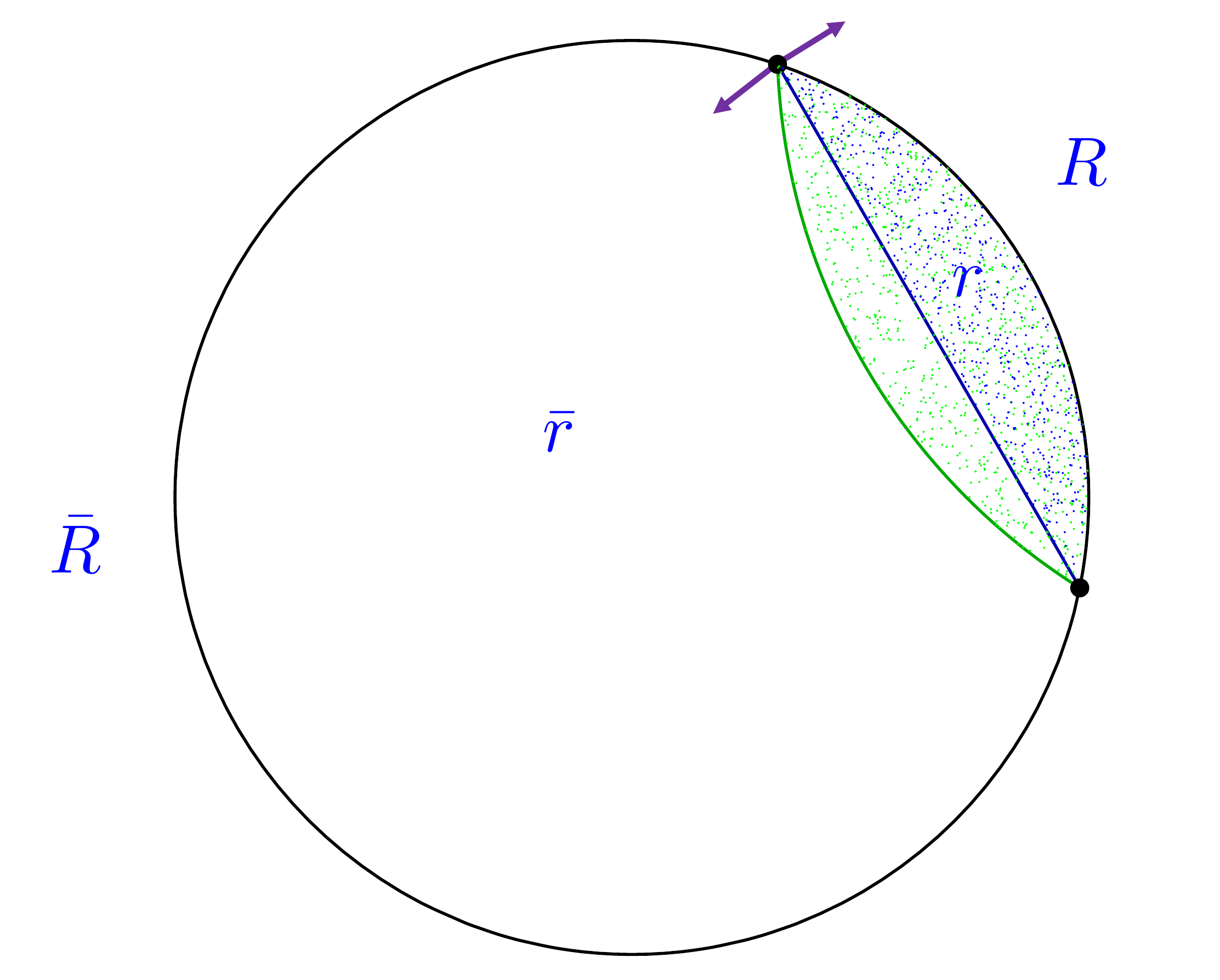}
\captionsetup{width=0.9\textwidth}
\caption{A schematic illustration of the relationships between the regions described in the text, on a timeslice which we can take for simplicity to be at the moment of time symmetry in the gravity-side geometry.  We show both the $Z_2$ symmetric case where the interval $R$ and its complement $\bar R$ are of the same size (left panel) and the case where they are unequal (right panel).  The Ryu-Takayanagi surface is indicated in blue, and it separates the bulk effective field theory subregion $r$ from its complement $\bar r$.   The causal wedge of the smaller subregion is indicated in light green in the asymmetric case, indicating that it exceeds the entanglement wedge (indicated by blue hatchmarks).  This leads to the intersection between the CW of the smaller subregion and the EW of the complementary region discussed in the text. The purple arrows indicate the action of the bulk modular Hamiltonian \cite{Faulkner:2017vdd}\ on probe bulk fields near the RT surface near the boundary, projected to the pictured spatial slice.  In the left picture, similarly to asymptotically AdS spacetime, the RT surface intersects the boundary orthogonally, while generically in the deformed theories this is not the case, and the modular evolution in the bulk has a component that is orthogonal to the boundary.}
\label{fig:spatialslice}
\end{center}  
\end{figure}

Beyond asymptotic $\text{AdS}$ geometries, the causal wedge need not to be smaller than the entanglement wedge\footnote{There have been earlier works on subregions in the presence of finite boundaries. For instance, \cite{Sanches:2016sxy, Nomura:2016ikr} discusses a generalization of HRT in more generic spacetimes anchored to a holographic screen, analyzed in general relativity (without a specific dual formulation). We thank D.~Marolf for useful communication.}. In fact, this is common for the duals of $T\bar{T}$ deformed theories for a region $R$ which fills up less than half the space. 
We illustrate this schematically in Fig.~\ref{fig:spatialslice}.

In such a situation, the entanglement wedge of the complementary region $\bar R$ overlaps with the causal wedge of $R$. In this setting, we are still going to consider local operators which inherit their locality from the extrapolate dictionary, since in the bulk we have a standard quantum field theory.  The structure of the commutators between various subregions depends somewhat on the particular $T\bar T+\dots$ theory under consideration, and we will present detailed calculations on both sides of the duality for cutoff AdS/Poincar\'e and cutoff dS/dS, the cases with a well defined CW up through order $c^0$. But in this general discussion, we will also note more general behaviors that can arise (and which may be generic).      

\begin{figure}[h!]
\begin{center}  
\includegraphics[width=.9\textwidth]{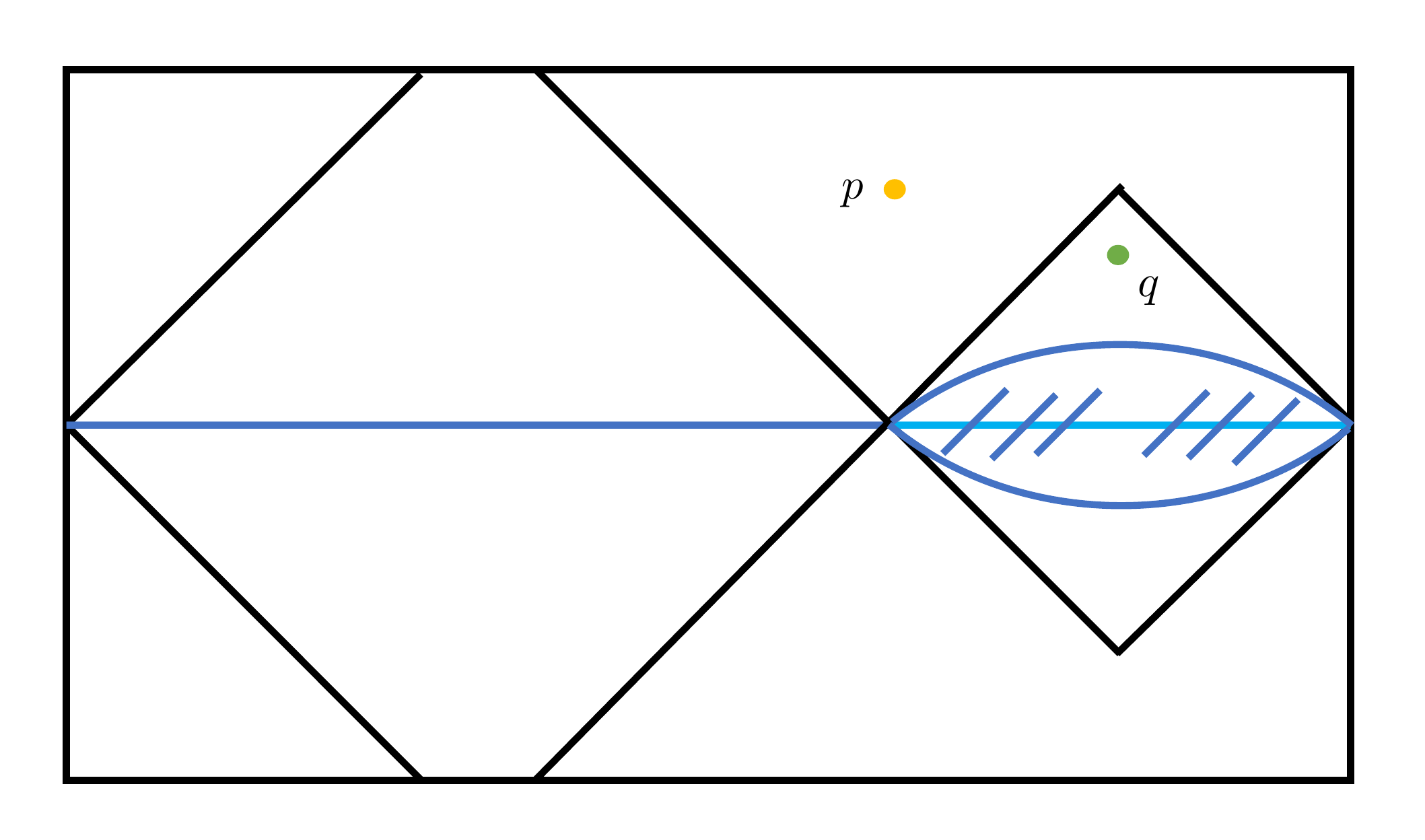}
\captionsetup{width=0.9\textwidth}
\caption{The regions $D[R]$ and $D[\bar R]$ for an asymmetric division of the system in the case of a boundary $dS_2$ spacetime, whose Penrose diagram we depict here.  The region $E[R]$ for $R$ the smaller subregion is indicated by the dark blue hatchmarks. Operators at the point $p$ can be causally influenced from either region and are not in EW[R] or EW$[{\bar R}]$; this is familiar in asymptotic AdS/CFT.
The point $q$ illustrates a point in $D[R]$ which is not in $E[R]$.  Operators at that point do not commute with those in the algebra ${\cal A}_{\bar R_s}$ corresponding to the complementary subsystem to $R$.  We note that this point $q$ on the boundary is also not part of the EW of either $R$ or $\bar R$, since it can be reached causally from both.}
\label{fig:timelikeboundarydS}
\end{center}  
\end{figure}

The causal wedge being larger than the entanglement wedge implies that there are $x \in D[R]$ such that $O(x) \not \in {\cal A}_{R_s}$ while for local unitary theories we are used to identifying the algebra of operators in $R$ with that of $D[R]$. 
We can quantify this effect by considering the largest region $E[R]$ such that $\forall x \in E[R], O(x) \in {\cal A}_{R_s}$. In our holographic setting, one can characterize this region explicitly. We depict that in Fig.~\ref{fig:timelikeboundarydS}. To contrast this with the standard case, it is useful to recall that in local field theories there are operators (e.g. on point $p$ in Fig.~\ref{fig:timelikeboundarydS}) that are not in the algebra of $R$ or $\bar R$. This point can be causally influenced  from either region. In the bulk dual, these are points that are not included in EW$[R]$ or EW$[\bar R]$. In our case, however, a point $q$ is completely inside the domain of dependence $D[R]$, but cannot be reconstructed from either ${\cal A}_{R_s}$ or ${\cal A}_{\bar R_s}$.

This has implications for the modular flow near the boundary of the region.  In QFT's, it is standard for the modular Hamiltonian to be locally Rindler when acting on operators near the boundary of the region $\partial R$, that is $[K_R,O(x \sim \partial R)]=\frac{d}{d\xi} O(x)$, with $\xi$ the boost generator. From the previous discussion, we don't expect this to be true anymore: for example, for $R$ less than half the space it is clear that this local behavior would quickly bring the operator outside of $E[R]$. One can also get some guidance into this using the bulk: in the bulk, the modular hamiltonian acts locally near the RT surface, which implies that it would act locally in the bulk near $\partial R$.
In the asymptotically AdS case, the RT surface hits the boundary orthogonally, implying that the bulk modular flow near the RT surface near the boundary acts within $R$ as a locally Rindler boost.  However, the generator of local bulk boosts won't be parallel to the boundary in the deformed theories, which implies that even close to $\partial R$, the modular hamiltonian can't act locally in the boundary. Because it acts locally in the bulk, we have that $[K_R,O(x \sim \partial R)]=\frac{d}{d\xi_\text{bulk}} O(x)$.  Given the presence of the holographic direction, this suggests than in addition to contributions from the stress tensor, which generate $\partial_{x_\text{bdy}}$, we will also get contributions from $T \bar{T}$ which generate $\partial_{\lambda} \sim \partial_{r_\text{bulk}}$.  This would be interesting to explore further in the future. 

The EW for the complementary region $\bar R$ covers the rest of the bulk; the union of the entanglement wedge of a region and its complement give the whole algebra ${\cal A}_{R_s} \cup {\cal A}_{\bar R_s}={\cal A}_\text{GFF,CFT}$. These entanglement wedge algebras, even if they are non-local in general, preserve the characteristic property of algebras of subregions: 
\be\label{eq:comm1}
[{\cal A}_{R_s} ,{\cal A}_{\bar R_s}]=0\,.
\ee
This, combined with the previous point, implies that one can get nonzero commutators between elements of ${\cal A}_{\bar R_s}$ and elements of ${\cal A}_{D[R]}$:
\be\label{crosscommutator}
[{\cal A}_{\bar R_s}, {\cal A}_{D[R]}]\ne 0\,.
\ee
That is, in the bulk the CW of a region can overlap with the EW of the complementary region! 
This happens even in cases where the CW makes sense, such as the dS boundary example of Fig.~\ref{fig:timelikeboundarydS}, or the AdS/Poincar\'e case in Fig.~\ref{fig1} below. We note that in this case it is possible to act with a unitary in $D[R]$, e.g. at point $q$ in Fig.~\ref{fig:timelikeboundarydS}, and change the entanglement entropy. This is because the unitary involves operators that are not inside the  algebra $\mc A_{R_s}$ that is relevant for the entropy of the region.

In the asymptotically AdS cases as previously mentioned, one generically encounters an EW[$ R$] 
that exceeds its causal wedge CW[$R$].  A clear example is the case with a region $R$ consisting of two disjoint intervals on the boundary, $R= R_1 \cup R_2$.  In those cases, the modular evolution does not stay within ${\cal A}_{R_1}$, but it does act within the total region $R$, i.e. it acts very nonlocally within $R$.  However, in the asymptotically AdS case, the new feature we just described of  $[{\cal A}_{\bar R_s}, {\cal A}_{D[R]}]\ne 0$ does not arise.       
This new effect clearly requires careful consideration, and we will focus on it in the next subsection.  First let us finish laying out the various possibilities for the commutator structure of our algebras in generality, just requiring bulk locality but no further specifications.

\subsubsection{Implications of bulk locality for boundary locality}

In this section, we are going to use the bulk to extract lessons about the locality of the algebras that we are considering, in order to understand better their support. This is possible because we just have a standard local theory in the bulk that allows us to do this. 

From the bulk point of view, we have that the union of the entanglement wedge of a region and its complement give the whole algebra ${\cal A}_{R_s} \cup {\cal A}_{\bar R_s}={\cal A}_\text{GFF,CFT}$. These entanglement wedge algebras, even if they are non-local in general, preserve the characteristic property (\ref{eq:comm1}) of algebras of subregions, even if $[{\cal A}_{\bar R_s},{\cal A}_{D[R]}] \not =0$ necessarily. 
Let us discuss the different inclusion relations between the entanglement and causal wedges, and their consequences.

When $\text{CW} \subset \text{EW}$ for both $R$, $\bar{R}$, we have a bulk region, denoted the causal shadow in \cite{Headrick:2014cta} that causal wedges can not reach. In other words, in these situations there are bulk operators that can't be reconstructed from either causal wedge:  ${\cal A}_{D[R]} \cup {\cal A}_{D[\bar{R}]} \subset {\cal A}_\text{GFF,CFT}$.

However, if  $\text{EW} \subset \text{CW}$ for both regions, given that the union of the entanglement wedges covers the whole bulk Cauchy slice, it is necessarily the case that there is causal intersection $C_{\cap} = \text{CW}[R]\cap \text{CW}[\bar{R}] \not = \emptyset $.\footnote{In flat space, this is true upon adding the point at infinity.} The points in $C_{\cap}$ can be reconstructed from either $\text{CW}[R]$ or $\text{CW}[\bar{R}]$ if HKLL continues to apply.  
A non-trivial $C_{\cap}$ implies that $[{\cal A}_{D[R]}, {\cal A}_{D[ {\bar R}] }] \not = 0$, or in other words one can go faster through the bulk than through the boundary and thus the boundary causal structure is not relevant.
We will not consider this situation in the bulk of this paper.  

Generically, the last possibility would be that $\text{EW}[R] \subset \text{CW}[R], \text{CW}[\bar{R}] \subset \text{EW}[\bar{R}]$ which could result in either a causal shadow or a causal intersection. If there is a causal shadow as in the case of a $\text{dS}$ boundary described in detail in \S\ref{sec:wcpiover2}, the boundary causal structure is respected.

All these general properties lead to the following conclusion: in the presence of causal intersections, the boundary causal structure becomes disrupted and the only meaningful dual of a subregion in this situation is the entanglement wedge. The entanglement wedge also has the nice property that it divides the bulk into two pieces which correspond respectively to two boundary subregions. We now return to the particularly subtle case (\ref{crosscommutator}) for the low energy fields, which is realized for Poincar\'e and dS boundaries.

\subsection{$\text{CW}[ R]\cap \text{EW}[\bar R]$ and the Hamiltonian and modular Hamiltonian evolutions}\label{sec:overlap}


Having introduced the algebras and some general implications, we now focus on the essential subtlety of the subregion dualities in the deformed theories identified above in (\ref{crosscommutator}).  
For this purpose, we will consider our main cases of interest: (A)dS/dS and AdS/Poincar\'e, for which the propagation is fastest on the boundary.  The essential novelty is that   $[{\cal A}_{\bar R_s}, {\cal A}_{D[{ R}]}]\not=0$ for finite volume when $R$ is smaller than half of the size of the system.  That is, the causal wedge of $ R$ intersects the entanglement wedge of $\bar R$.

We can express this succinctly in the following way.  Given EW reconstruction of EW[$\bar R$] and HKLL reconstruction of CW[$R$], we have the formula for the reconstruction of some bulk operator $\Phi(X_r)$ in this region
\bea\label{Twoways}
\Phi (X \in {\rm EW}[\bar R] \cap {\rm CW}[ R])& =& \int_{\bar R} d x_{\bar R}\int d s\,f_{\Delta ,s}^{\bar R}\left(X,x_{\bar R} \right)\rho_{\bar R}^{ - is/2\pi }O\left( x_{\bar R} \right)\rho_{\bar R}^{is/2\pi }\nonumber\\
&= &\int_{D[ R]} {\tilde f} \left( X, x_{D[R]} \right){U^\dag }\left( {{t_{D[ R]}},{t_{ R}}} \right)O\left( x_{R} \right)U\left( {t_{D[R]}, t_{R}} \right)
\eea

We used \cite{Faulkner:2017vdd} in the top line, and our version of HKLL \cite{Hamilton:2005ju}\ in the bottom line.  We stress that this notation for the bulk operators includes both the fields $\phi$ and their conjugate momenta $\pi_\phi$, and as above we focus on their leading behavior as free fields without backreaction at order $c^0$.  
In the first line, we formulate the bulk operator via
modular evolution; in the second line, we have used Hamiltonian evolution to write it in terms of an operator at $t=t_ R$, with $U=T e^{-i\int H}$ in terms of the Hamiltonian $H$.  

Consider formulating a bulk field $\phi$ in the overlap region using the top line, and its conjugate $\pi_\phi$ at the same point using the bottom line of (\ref{Twoways}).  
First, we use the fact that  $[O(x_R), O(x_{\bar R})]=0$.  Thus in order to obtain the required nonzero canonical commutator in the bulk, $[\pi(X), \phi(X')]=i \delta(X-X')$, it must happen that
\be\label{rhoU}
[\rho_{\bar R}, O(x_ R)]\ne 0\,,\; \quad [U_{D[ R]}, O(x_{\bar R})]\ne 0\,, \quad {\text{or}} \quad [\rho_{\bar R}, U_{D[R]}]\ne 0 \,,
\ee
where we recall that $R$ is a region with size less than half of the system (a finite interval in AdS/Poincar\'e).
Both $\rho_{\bar R}$ and $U$ are potentially subtle in our deformed theories.  
As noted above, we cannot a priori assume that the Hilbert space behaves as in local quantum field theory, which in the presence of a lattice regulator factorizes along the nominal boundary subregions $R, \bar R$.  A priori this raises the possibility that there is no such division of the system in the deformed theory, so that $\rho_{\bar R}$ may act beyond the putative region $\bar R$. 
Indeed, the bulk modular flow heads into the bulk rather than staying at the boundary in our cases of interest (see Figs.~\ref{fig:spatialslice} and \ref{fig1}), because in the cutoff theories the RT surface does not intersect the boundary orthogonally as discussed above.  

However, by construction
$K_{\text{bulk}, \bar r}$ does not act in EW$[R]$, including at the boundary $R$ where the local operators $O(x_{R})$ reside.
This we inferred above already in writing $[{\cal A}_{R_s}, {\cal A}_{\bar R_s}]=0$.  So it appears that the first commutator in (\ref{rhoU}) vanishes.
In general we also cannot assume that the Hamiltonian evolution acts within $D[ R]$. But at least within low energy effective theory of the vacuum in the boundary dS and Poincar\'e cases of interest, here we do have causal Hamiltonian evolution, up to and including order $c^0$ effects on probe bulk fields. 
This suggests that the second commuator in (\ref{rhoU}) vanishes.  We stress that non-locality and acausality are not identical, as in string theory, and the fact that propagation is fastest along the boundary in the present examples is consistent with causality.   This leaves the third commutator in (\ref{rhoU}),
\be\label{eq:novel2}
[\rho_{\bar R}, U_{D[R]}]\ne 0 \,.
\ee   
In any case, regardless of whether the nonlocality is to do with $\rho$, $U$ or both, we can package the novelty in the need for a nontrivial commutator (\ref{eq:novel2}).  

In the bulk of this paper, we provide new computations of entanglement entropy, and of the behavior of the stress energy on the Euclidean geometries relevant for the R\'enyi entropies and for the density matrix itself.  The ultimate interpretation of the intersection $\text{CW}[R]\cap \text{EW}[ \bar R]$ must respect all these results.  We will return to discuss this briefly in the conclusions.

\subsection{Case study: an interval in cutoff AdS}

The simplest example of $ \text{EW }> \text{CW}$  is cutoff AdS. Recalling the discussion in \S\ref{subsec:bSSAgr}, the entangling surface is the same as in vacuum AdS, in Poincar\'e coordinates: 
\begin{equation}
z^2+x^2=L^2+z_c^2, ~~ z \in (z_c, \sqrt{L^2+z_c^2} )\,.
\end{equation}
The difference is that it is anchored to $z=z_c$. On the other hand, the causal wedge  is delimited by a semicircle centered at $z=z_c$: 
\begin{equation}
(z-z_c)^2+x^2=L^2 ,~~  z \in (z_c, L+z_c )\,.
\end{equation}
We see that the CW contains the entanglement wedge because the causal wedge reaches further, $\sqrt{L^2+z_c^2}<L+z_c$.

\begin{figure}[htbp]
  \centering
  \includegraphics[width=0.6\textwidth]{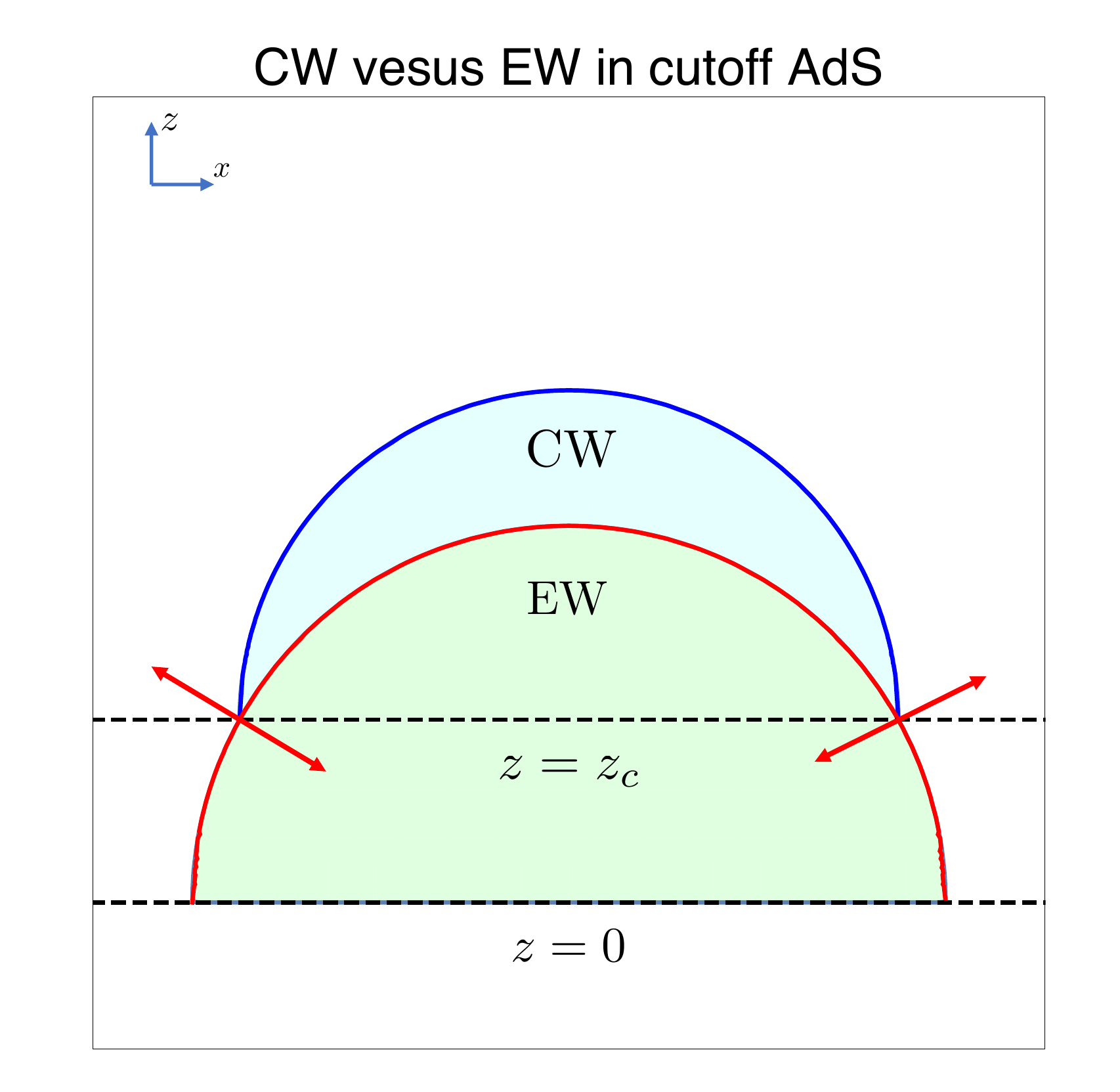}
  \caption{\label{fig1} CW vesus EW in cutoff AdS. When we move the boundary from $z=0$ to $z=z_c$, CW becomes larger than EW. The modular Hamiltonian evolution near the boundary is given by the inward pointing red arrow for $R$, and points outwardds for $\bar R$.}
\end{figure}

In this situation, it is easy to get the EW and $E[R]$ (defined around Fig.~\ref{fig:timelikeboundarydS}). One can get the boundaries of $E[R]$ by shooting light-rays from the RT surface towards the boundary. Since the lightrays just follow the equation $t=|\sqrt{L^2+z_c^2}-r|$, the region $E[R]$ is bounded by
\begin{equation}
t= \pm ( \sqrt{L^2+z_c^2}-\sqrt{x^2+z_c^2}  )\,.
\end{equation}
The description of this phenomena is shown in Fig.~\ref{fig1}. In this case, the proof that works for asymptotic AdS, saying that EW should always be not smaller than CW \cite{Hubeny:2012wa,Wall:2012uf,Headrick:2014cta}, does not apply for cutoff AdS. \footnote{The proof based on the inequality of expansions of surfaces in GR \cite{Hubeny:2012wa,Wall:2012uf}, does not hold for cutoff surfaces in general. The reason is that there might exist causal wedges with negative expansions, which violate the assumption we made for asymptotic AdS. We thank A.~Wall for discussions.}

There are no caustics, so we expect that the algebra of the entanglement wedge is the same as the algebra of $E[R]$: ${\cal A}_{R_s}={\cal A}_{E[R]} \subset {\cal A}_{D[R]}$. It would be nice to understand this in terms of the modular flow, but we leave this for future work. 
Note that in this case, the analogue of the causal wedge if we replace $D[R]$ with $E[R]$ is the same as the entanglement wedge. 
In this situation one can't travel faster through the bulk than through the boundary, since outgoing light rays thrown through the boundary will never make it back the boundary unless they are parallel to it; equivalently, we have that $[{\cal A}_{D[R]},{\cal A}_{D[\bar{R}]}]=0$.  

Given these results, let us return to the violation of boosted SSA above in \S\ref{sec:bSSA}. Along the lines of what we just discussed, we can compute the EW associated to the boosted intervals $A, B$, in Fig.~\ref{fig:bSSA}. This shows that $\text{EW}[A] \not \subset E[A \cup B]$, and hence we do not have the inclusion property for the corresponding algebras, $\mc A_{A_s} \,\not \subseteq\, \mc A_{(A \cup B)_s}$. This explains why the boosted SSA does not have to hold: we do not have (\ref{eq:locality2}), so we cannot use the monotonicity property (\ref{eq:monotonicityrel}). In contrast, $\text{CW}[A]  \subset \text{CW}[A \cup B]$, and hence $\mc A_{D[A]} \,\subseteq\, \mc A_{D[A \cup B]}$ in the low energy theory of the bulk gravitational fields. For these algebras we would have a violation of boosted SSA. For this reason, it is more natural to associate the algebra of local operators with $\mc A_{R_s}$ and not with $\mc A_{D[R]}$.

A closely related aspect is that a unitary transformation localized inside $D[R]$ can change the entropy. From the holographic dual, this is a simple consequence of $\text{EW}[R] \subset \text{CW}[R]$. A perturbation in CW$[R]$ but not in EW$[R]$ can causally influence the extremal surface and hence change the entropy.

\subsection{Case study: dS/dS $w_c= \frac{\pi\ell}{2}$}\label{sec:wcpiover2}

In this section, we make a more detailed study of subregions and entropy for a full throat of dS/dS, with $w_c=\pi\ell/2$, dual to a trajectory $T\bar T+\Lambda_2+\text{curvature}\dots$ as described in \S\ref{sec:setup}.  We will provide both gravity-side and field theory side calculations, finding explicit agreement.  First, in the next section we introduce the geometry of the CW and EW, both for more general $w_c$ and for the special case $w_c=\pi\ell/2$.  Then in the following two sections we analyze the entropy and the structure of the density matrix at large $c$ by analyzing in two ways the equations for the stress energy (\ref{Teqs}).   

\subsubsection{The subregions and algebras}

\begin{figure}[htbp]
  \centering
  \includegraphics[width=0.55\textwidth]{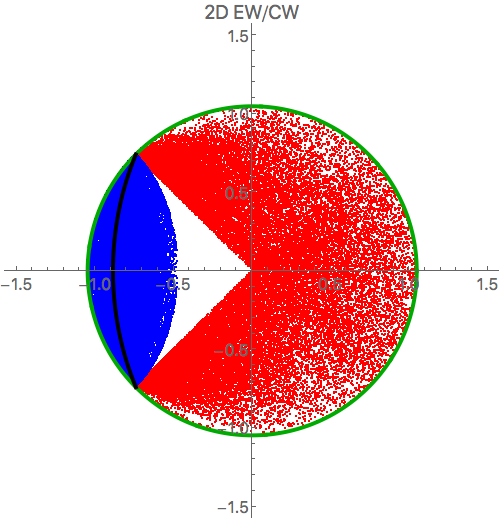}
  \caption{\label{fig2} A more general example. We consider region $|R|=\ell \pi/2$ and $w_c=\pi \ell /3$. We define $x=w/\ell \cos\phi$ and $y=w/\ell \sin \phi$ in the dS/dS slicing. Red region: causal wedge in $\bar{R}$; Blue region: causal wedge of $R$; Black curve: extremal surface (entanglement wedge); Green curve: the boundary $w=w_c$. The causal wedges in the plot are made by sampling numerically several initial conditions for geodesics starting from the domain of dependence.}
\end{figure}

For general $w_c$, we find the phenomenon noted above that the CW can exceed the EW, that the CW of one region can intersect the EW of the complementary one, and that there can be causal shadows.  This is depicted in an example
via numerically sampling the solution of the geodesic equations passing through domain of dependence in 
Fig.~\ref{fig2}. 


We can understand the (A)dS/dS causal shadow very simply geometrically, as follows.  Consider the boundary $\text{dS}_2$ depicted above in Fig.~\ref{fig:timelikeboundarydS}.   
If the whole diamond is contained in the dS patch, the causal wedge is just the trivial bulk extension. However, if the diamond gets truncated into a hexagon at the spacelike future boundary of the $\text{dS}_2$ (as happens for $\bar R>\pi\ell$ at $t=0$), the causal wedge is the union of causal wedges that derive from the union of diamonds which live inside $D[\bar R]$.
Since there is no entanglement shadow, we will focus on the entropies and properties of modular Hamiltonian evolution below.

In the particular case $w_c=\pi\ell/2$, these features become extreme.  The extremal surface, which is always a great circle, is now located along the boundary for any unequal division of the system: $R\ne \pi\ell$.  As a result, the corresponding $E[R<\pi\ell]$ only contains $R$ itself.  That in turn implies that the modular Hamiltonian acts trivially on the operators in $R<\pi\ell$.    
Conversely, the entanglement wedge of a region $R>\pi\ell$ covers the entire bulk space $r$ at time $t=0$. The modular flow points straight inward from the RT surface in this case.  These features were also noted in the earlier work \cite{Pastawski:2016qrs}.
Below in \S\ref{sec:QEC}\ we will comment on the implications for quantum error correction and redundancy of the encoding of the bulk into the operator algebras of the dual deformed CFT.






\subsubsection{Entanglement and the modular Hamiltonian in the 2d dual via the deformed stress energy}

In this section we will consider in detail the density matrix and entanglement structure of 
the dS/dS $w_c=\pi\ell/2$ theory.
Dividing the boundary at $\tau=0$ along an interval of proper size $R$ (see Fig.~\ref{fig:spatialslice}), we distinguish three main cases. If $R$ is half the space, i.e. $R=\pi\ell=\bar R$, there is additional symmmetry in the problem:  the equal division into $R$ and $\bar R$ implies that $D[R]$ and $D[\bar R]$ are complementary static patches in the boundary $\text{dS}_2$, preserving the corresponding time translation symmetry.  In this special case, $\text{CW}=\text{EW}$.  The calculation of the entanglement entropy was completed on the deformed CFT side in \cite{Gorbenko:2018oov}\ (for any $w_c\le \pi\ell/2$), matching the bulk entropy obtained via Ryu-Takayanagi.  
We will shortly generalize this to obtain more information from the R\'enyi entropies in this case,
showing that the system is maximally mixed for $R$ less than or equal to half the system.  That indicates that the modular flow is trivial, since the density matrix is proportional to the identity.
In fact, as we will see shortly, this example exhibits the effect already at large $c$, and we will show that it persists to the next order under a small deformation. 

In order to analyze this, we will require the stress energy tensor satisfying (\ref{Teqs}) in the appropriate geometry.  
Let us start with the pac man diagram for $\rho_R$, for arbitrary interval $R$.  This is two
hemispheres, sewn together along $\bar R$ and with independent Dirichlet boundary conditions for the fields at $R$ on each half.  We will be interested in the solutions for $T_{ab}$ both on this pac man geometry for $\rho_R$ itself, as well as the replicated geometry obtained by stitching together $n$ copies of this in order to compute $\tr(\rho_R^n)$ and the R\'enyi entropies.

The euclidean metric is
\be\label{methem2}
ds^2=r^2(d{\theta}^2 + \sin^2{\theta} d{\phi}^2)\,,
\ee
with $\phi$ the analytic continuation of the $\text{dS}_2$ static patch time.
When the system is divided in half, $R=\pi\ell$, the region $R$ is the locus $\phi=0, 0\le\theta\le\pi$, and similarly for a smaller region $R<\pi\ell$ which covers a smaller range of $\theta$.  For $R$ greater than half, the region extends to include both the above and the locus $\phi=\pi, 0\le\theta\le (R-\pi\ell)/\pi\ell$.

Using the Christoffel symbols,
\be\label{Christoff2}
\Gamma^\theta_{\phi\phi}=-\cos\theta\sin\theta, ~~~~ \Gamma^\theta_{\phi\theta}=\cot\theta
\ee
we find that the equations (\ref{Teqs}) become
\bea\label{TeqsS3}
0 &=& \partial_\phi T^\phi_\phi+\partial_\theta(\sin^2\theta T^\phi_\theta)+  T^\phi_\theta \cos\theta\sin\theta \nonumber\\
0 &=& \partial_\theta T^\theta_\theta+\partial_\phi(T^\phi_\theta)+  (T^{\theta}_{\theta}-T^{\phi}_{\phi} )\cot{\theta} \nonumber\\
T^{\theta}_{\theta}+T^{\phi}_{\phi} &=& -\frac{ c {\mc R^{(2)}}}{24\pi}+\frac{c_2}{\pi\lambda}+\pi\lambda(T^{\theta}_{\theta} T^{\phi}_{\phi}-(T^{\phi}_{\theta})^2\sin^2{\theta}) 
\eea 
with appropriate boundary conditions for a given application.  For a calculation of the R\'enyi entropies, one needs a boundary condition which specifies that the theory is in the desired state, e.g. the vacuum.  This condition is straightforward in the $R=\pi\ell$ case with extra symmetry, as well as in the AdS/Poincar\'e case analyzed above in \S\ref{sec:bSSA}, but we have not implemented it in general. For the pac man itself, we may consider different entries in the density matrix, which correspond to different choices of boundary conditions.   

In the next two subsections, we will apply these equations determining the large-$c$ stress energy tensor in two ways.  First, we will calculate the zeroth and first R\'enyi entropies in the symmetry case, $R=\pi\ell$.  
Then, we will analyze the behavior of solutions on the pac man, with different choices of $R$.  In both cases, we find that the independent dual deformed-CFT calculation matches the gravity side.

\subsubsection{R\'enyi entropies and maximal mixing}

In \S\ref{subsec:DS}\ we revisited 
the calculation of the R\'enyi entropies for the half and half division of the system, $R=\pi\ell$, in the $T \bar T $ and $T\bar T+\Lambda_2$ theories on $\text{dS}_2$  
in \cite{Donnelly:2018bef}\cite{Gorbenko:2018oov}.   We will now use those results to compute the R\'enyi entropy.

\subsubsection*{R\'enyi entropies and maximal mixing}

We need to evaluate
\be\label{eq:Renyi}
S_n= \frac{1}{1-n}\,\log\,\tr\,\rho^n =\frac{1}{1-n}\,\log\frac{Z_n}{Z_1^n}\,,
\ee
with $Z_n$ the partition function on the replicated manifold. This is determined by the VEV of the stress tensor trace, via (\ref{eq:ZnTrTSn})
\be\label{eq:Zn}
\partial_r\,\log Z_n= -2 \pi n r\,\int_0^\pi d\theta\,\sin\theta\,(T^\theta_\theta+T^\phi_\phi)\,.
\ee
As a check, for $n=1$ the angular integral is trivial, and we reproduce the result for the dS/dS sphere partition function \cite{Gorbenko:2018oov} 
\be
\partial_r\,\log Z_1(r) =-\frac{8 r}{\lambda}\left(1-\sqrt{\eta+\frac{c\lambda}{12 r^2}} \right)\,.
\ee

The simplest R\'enyi entropy to evaluate is for $n=0$, and we will use this to establish that the density matrix is maximally mixed at large $c$ and for $\text{dS}_2$ at $w_c=\frac{\pi}{2}\ell$. In this case, we will also be able to obtain the entropies for smaller intervals, and match with the gravity side result.

We have
\be
S_0 = \log \text{tr}_R \mathbf 1 = \log \text{dim} \mathcal H_R\, = \log Z_0\,.
\ee
At $n=0$, our expressions in \S\ref{subsec:DS} for the stress energy tensor give, up to terms that vanish as $\delta \to 0$,
\be
\partial_r\,\log Z_0= -2 \pi  r\,\int_0^\pi d\theta\,\sin\theta\,(n\, \tr T)\Big|_{n=0} = \frac{2\pi }{\lambda} \sqrt{\frac{c\lambda}{12 }}\,.
\ee
If we fix the integration constant via the gravity side, so that the partition function vanishes at $r=0$, we obtain
\be
S_0(r) = \frac{2\pi}{\lambda} \sqrt{\frac{c\lambda}{12 }} \,r\,.
\ee
We will choose the integration constant this way in general in this section.  It will interesting to analyze the integration constant directly in the deformed QFT, but we leave that for future work.  

Let us compare this with the entanglement entropy,
\be
S_1(r) = \frac{c}{3} \text{arcsin}(\text{h})\,\frac{r}{\sqrt{\frac{c\lambda}{12 }}}\,.
\ee
For $\eta=1$, corresponding to bulk AdS, the entropy is the $\text{arcsinh}$, and $S_1(r) < S_0(r)$ for all $r$. So the state is not maximally mixed. However, for $\eta=-1$ (bulk dS), we have
\be
S_0(r) = S_1(r)=\frac{\pi c}{6} \;\;\text{for}\;\;r=\sqrt{\frac{c\lambda}{12 }}\,.
\ee
In gravity language, this corresponds to the central slice $w_c= \frac{\pi}{2} \ell$.

Recall that $S_1 \le S_0$ always, and $S_1 = S_0$ if and only if the state is maximally mixed,
\be\label{eq:maximal1}
\rho_R = \frac{\mathbf 1_R}{\text{dim} \mathcal H_R}\,.
\ee
So we learn that for bulk dS at the central slice, the density matrix corresponding to tracing over half of the system is maximally mixed. This also implies that the entanglement spectrum is flat, and all the R\'enyi entropies have to be the same. This is quite different from what happens in the AdS case, where the R\'enyi entropies for $n>1$ are different (but we also do not have maximal mixing, so there is no contradiction).

\subsubsection*{Perturbation away from large $c$}

Moreover, we can see in general that this statement of maximal mixing will survive a small deformation, at least at the first order in the deformation.  First, to make sure to keep track of the normalization, let us introduce the following notation and conventions.   Note that the order $c$ and full density matrices will in general have different rank since we will get some correction $\delta S_1 = S_1-S_1^{(0)}$ to the (finite) entropy from stringy and/or quantum effects.    (Recall that the entropy is finite because we truncate to real energy levels in the dressed theory.)  
We can embed the smaller of these two matrices in the larger one by adding at least $e^{|\delta S_1|}$ rows and columns with entries $0$ to the one with the smaller entropy. 
The leading order density matrix is diagonal, with a block of the form $e^{-S_1^{(0)}}{\bf 1}_{e^{S_1^{(0)}}\times e^{S_1^{(0)}}}$.  The full density matrix $\rho_1$ has a block of dimension $e^{S_1}\times e^{S_1}$.   So in this notation, they are matrices of the same dimension but different rank.   

Then for the perturbed R\'enyi entropies, we can write:
\be\label{pertK1}
\delta \tr(\rho_1^n) = n \,\tr ((\rho_1^{(0)})^{n-1}\delta\rho_1)   \,.
\ee
For $n=1$ this must vanish, since the normalized density matrix has $\tr( \delta\rho_1)=0$.  In fact, for higher $n$ any potential contribution to (\ref{pertK1}) would be second order in the deformation:  the matrix $\rho_1^{(0)}$ is proportional to the identity, and the difference in its rank from the full rank is part of the deformation.  Since we have one power of $\delta\rho_1$ in (\ref{pertK1}) already, that difference in rank would arise only at second order.  Thus we can ignore the difference in rank, and use that $(\rho_1^{(0)})^n\propto {\mathbf 1}$ to conclude that (\ref{pertK1}) vanishes for all $n$ at least to first order in the deformation.    
This means that the flat entanglement spectrum persists at least to first order in the correction.  This result depended crucially on the flat entanglement spectrum  at leading order.

\subsubsection*{Entropies for non-symmetric subsystems and random pure states}


For dS/dS with $w_c=\pi\ell/2$, and division into half ($R=\pi\ell$), we have just found a maximally mixed state (\ref{eq:maximal1}) in the large $c$ approximation.
This means that the same should be true for smaller subsystems, since the density matrix is the identity in every basis of the Hilbert space. Physically, since every degree of freedom inside the subsystem is maximally entangled with something outside, doing a further partial trace will not change this.
So (\ref{eq:maximal1}) (and as a corollary the flat entanglement spectrum) should also hold for smaller systems with $R<\pi\ell$.

The large-$c$ and perturbative maximal mixing agrees with the gravity-side result. 
The extremal curve collapses to the boundary, consistent with a trivial modular Hamiltonian for smaller subregions, $R<\pi\ell$. So in this case we also match the boundary and bulk modular Hamiltonians.  
From the gravity side, the entropy is
\be\label{eq:Ssubsystem}
S(\Delta \theta) = \frac{\pi c}{6}\,\frac{\Delta \theta}{\pi}
\ee
and $\Delta \theta$ is the angular size of the subsystem. This is just the length of the geodesic inside the great circle at $w_c = \frac{\pi}{2}\ell$.

For a subsystem that is larger than half of the system size, we use that the global state is pure, and hence its entropy should equal the entropy of the complement, for which we established the maximal mixing. Combining both results obtains
\be\label{eq:random}
S(A) = \frac{c}{6}\, \left \lbrace \begin{matrix} \Delta \theta\;&,&\;\Delta \theta < \pi \\ 2\pi -\Delta \theta\;&,&\; \Delta \theta >\pi \end{matrix}\right.\,.
\ee
This also follows simply from the length of geodesics on the sphere.

For a small subsystem this gives a volume law; so small subsystems behave as (approximately) maximally mixed. However, once the size of the subsystem reaches half of the system size, the density matrix contains enough information to realize that the complete state is pure. These properties identify the de Sitter vacuum state as a random pure state~\cite{Page:1993df}. This aspect was already noted in~\cite{Pastawski:2016qrs}. A similar behavior is observed in the holographic mutual information, which changes from zero to a nonzero value as more than half of the system is probed. It would be interesting to understand in more detail how the random pure state structure arises from the $T \bar T+ \Lambda_2$ flow.

\subsubsection{Properties of the modular Hamiltonian from stress energy on the pac man}\label{sec:dScharspacman}

In order to understand the behavior of the modular Hamiltonian for $R\ne\pi\ell$, we must exhibit 
its novel action on $D[\bar R]$ for $R>\pi\ell$ as well as its reduced action for $R<\pi\ell$.
In the previous section, we derived the latter statement via a calculation of the maximally mixed density matrix for $R\le\pi\ell$.  In this section, we will study the 2d theory's stress energy directly on the pac man geometry relevant for computing the density matrix (or equivalently $K$) itself, finding new behaviors for $R\ne\pi\ell$.

In this approach, we will study our differential equations using the method of characteristics, focusing on the present example of dS/dS at $w_c=\pi\ell/2$.   First, let us analyze this in more generality to see the effect of the curvature and the $\Lambda_2\sim (1-\eta)/\pi\lambda$ term in the trace flow equation.  Following the method briefly reviewed in Appendix \ref{sec:Charappendix}, we find a system of equations of the form
\be\label{Consmatrix}
\partial_\phi \vec U + B \partial_\theta\vec U=\vec C\,.
\ee
The eigenvalues of $B$ which determine the reality of the characteristics are
\bea\label{Revalues}
\lambda_\pm &=&  \frac{1}{(1-\tilde\lambda u_1)}\left(\tilde\lambda u_2(\cos(2\theta)-1)) \pm \sin\theta\sqrt{c_2-1-c\tilde\lambda R/24\pi} \right)\nonumber\\
&=& \frac{1}{(1-\tilde\lambda u_1)}\left(\tilde\lambda u_2(\cos(2\theta)-1)) \pm \sin\theta\sqrt{-\eta-c\tilde\lambda R/24\pi} \right)\nonumber\\
&=&\frac{d\theta}{d\phi}\,,
\eea
where $\tilde\lambda=\pi\lambda$, $u_1=T^\phi_\phi$ and $u_2=T^\phi_\theta$.  In the last step we used equation (\ref{charsgen}).  
This formula cleanly distinguishes different cases. For $\eta=1$ (the AdS/dS case) it is elliptic: $\lambda_\pm$ are complex.  For the dS/dS theory, in our special case $w_c=\pi\ell/2$  of interest in this section, it is parabolic and we have real characteristics.  For dS/dS with $w_c<\pi\ell/2$ cutoff, however, it is elliptic.  As in our discussion of the causal wedge, the boundary dS case is generally closer to the original AdS/CFT duality than the boundary cylinder case; in comparing our results in this section with those in Appendix \ref{sec:Charappendix}, we see another manifestation of that.    

Let us continue to focus on the $w_c=\pi\ell/2$ case.  As explained in the previous subsection, we have an effect on $K$ that arises first at large $c$, where these equations for $T_{ab}$ apply.
We want to understand the effect on the modular Hamiltonian of changing $R$ from half the system $R=\pi\ell$ to above or below this.  This is a question only about the locus $A$ on which we put boundary conditions:  in all cases we are considering the same set of PDEs on the same space (here a 2-sphere).  
One gets a well posed initial value problem if the set of characteristics have the property that they each cross $A$ precisely once.  If some characteristics cross $A$ more than once, this can lead to singularities or discontinuities in the solution.  If they don't all cross $A$, then the problem is underdetermined.

From this, we can now explain the essential difference between $R=\pi\ell$, $R>\pi\ell$, and $R<\pi\ell$ in this language.
To do that in a very simple way way, let us consider in particular the set of characteristics that we get by setting $u_2=T^\phi_\theta=0$.  These are very simple, going around the sphere in the $\phi$ direction, along curves of constant $\theta$ (since $u_2=0$, they satisfy $\lambda_\pm=\frac{d\theta}{d\phi} = 0$ as easily seen from (\ref{Revalues}) recalling that the square root term is zero).

First, consider region $R=\pi\ell/2$.  In this case, the characteristics cross $R$ precisely once each.  So putting the boundary condition on $R$ in this case constitutes good initial data.   Now we see exactly what happens in the $R\ne \pi\ell$ cases.  If we extend the region to $R>\pi\ell$, some of these characteristics cross twice.  This is the case that corresponds to there being discontinuities arising from the crossing of different characteristics.  To see that here, consider boundary conditions as above on the original half-space locus, but on the extra portion of $R$ we can put different boundary conditions.  The characteristics that emanate from that additional region within $R>\pi\ell$ generically inersect the first set which emanate from the original half space region.  

The crossing of characteristics is a harbinger of discontinuities in the solution, requiring additional data to define it.  This occurs precisely in the situation noted above in \S\ref{sec:overlap}\ where a nontrivial commutator (\ref{rhoU}) must arise.
The fact that this introduces discontinuities, requiring extra data to define patchwise solutions for $T_{ab}$, may be tantamount to extending the mouth of the pac man, indicating an action of the modular Hamiltonian beyond $R$.  However, this is a Euclidean calculation, and in general the crossing of the characteristics can occur away from the equatorial slice containing $R$ and $\bar R$ (although for some entries of the matrix, the crossing might occur on this locus).   

Conversely, if we consider $R<\pi\ell$, then the PDEs are underdetermined.    
In sum, these deformed-CFT side calculations fit with the gravity-side predictions for novel behavior of the modular Hamiltonian in our system when $R$ deviates from half the space.

\section{Comments on quantum error correction}\label{sec:QEC}

In the previous sections, we derived some features of subregion dualities and algebras for the nonlocal $T\bar T+\dots$ deformed theories which are holographically dual to patches of (A)dS, probing them with explicit calculations of entropy in key cases.   We will now make more explicit some implications for redundant encoding and quantum error correction, making contact with related literature. We will also discuss the potential of quantum simulation for quantum cosmology.  Let us begin by reviewing some of the basic ideas.

\subsection{Redundant encoding and quantum error correction}

The phenomenon of redundant encoding in holography is closely related to quantum error correction in quantum information theory --see for instance~\cite{Almheiri:2014lwa,Pastawski:2015qua,Dong:2016eik,Harlow:2016vwg}. In the language of quantum error correction, the effective field theory of gravity in the bulk is described by the code subspace, which corresponds to the low energy Hilbert space in the dual CFT. The whole Hilbert space in the CFT, whose qubits are called physical qubits, is used to redundantly encode quantum information, while qubits in the code subspace are called logical qubits. In this formalism, if one could recover the bulk points from a region $R$ in the boundary, then it equivalently says that the bulk point is protected from erasing Hilbert space in the region $\bar{R}$, while it is not protected from erasing the Hilbert space in region $R$. The error here is understood as the erasing error, and the modular flow or HKLL formula is understood as the recovery map from the boundary to the bulk. Thus, the language of (operator) quantum error correction codes is identified with redundant encoding of bulk points in the gravity side description of the system.

\begin{figure}[htbp]
  \centering
  \includegraphics[width=0.5\textwidth]{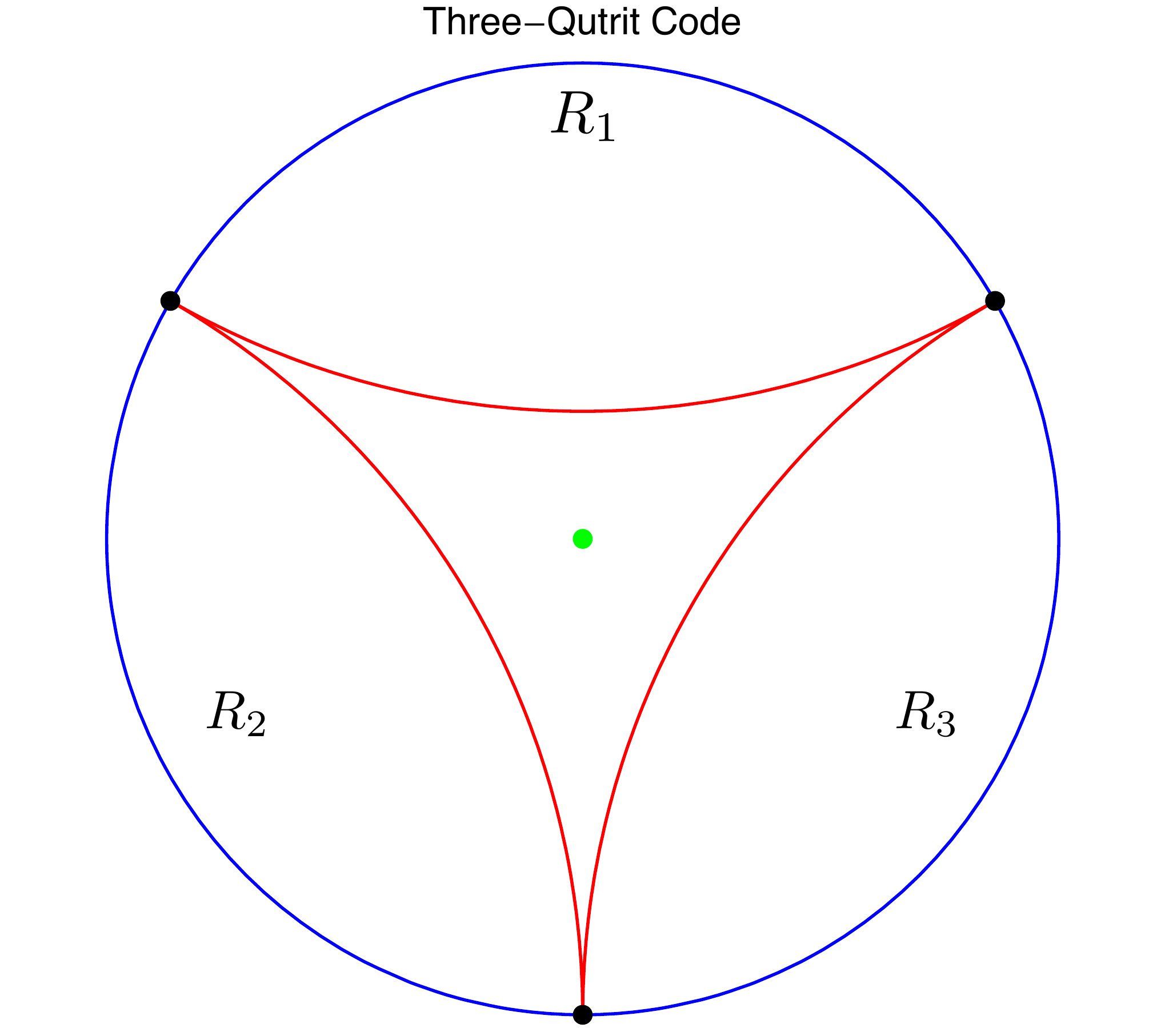}
  \caption{\label{fig3} The Three-Qutrit Code example.}
\end{figure}

One starting point of holographic error correction is from the commutator puzzle: considering a bulk operator $\Phi(X)$ where $X$ is at the center of pure AdS, and a boundary operator $\phi(x)$ where $x$ is an arbitrary boundary point on the same time slice, since $X$ and $x$ are space-like separated we expect
\begin{align}
\left[ {\Phi (X),\phi (x)} \right] = 0~.
\end{align}
However, this formula is only valid in the code subspace. In fact, if it holds as an operator equation for all points $x$ on the boundary, by Schur's lemma we see the operator could only be the identity, which is a contradiction since  we expect holographic duality should work for all operators on each side. A simple example of this phenomenon is the three-partition (see for instance, a recent review in \cite{Harlow:2018fse}). In Fig.~\ref{fig3}, the operator in the middle is protected by the quantum error correction code and cannot be recovered from any of the three boundaries $R_1, R_2, R_3$, but it could be recovered by any of two of these three regions, equivalently by any one of the three boundaries $\bar{R}_1, \bar{R}_2, \bar{R}_3$ by CW or EW reconstruction. 

A toy model  for this from quantum error correction theory is the Three-Qutrit Code, which has a 27 dimensional physical Hilbert space and a 3 dimensional code subspace. The commutator is only vanishing in the code subspace~\cite{Almheiri:2014lwa,Dong:2016eik,Harlow:2016vwg,Harlow:2018fse}. 

\subsection{The level of redundancy in our reconstructions}\label{sec:redundancy}

First, consider CW reconstruction based on the HKLL prescription which we generalized to our case in (\ref{HKLLgen}) and \S\ref{sec:HKLLappendix}. A clear feature of this in our geometries is the existence of the causal shadow. In EW reconstruction, in contrast, the entangling surface divides the bulk into two complementary regions, and any point in the bulk { at $t=0$} could either be recovered by the algebra ${\cal A}_{R_s}$ associated to region $R$, or by ${\cal A}_{\bar R_s}$ associated to region $\bar{R}$. This is called geometric complementarity \cite{Pastawski:2016qrs}. However, coming back to CW reconstruction, which is based on geodesics shooting from domain of dependence of $D[R]$, there can exist the shadow region in the bulk, $\text{CS}[R;\bar{R}]$, whose elements can neither be recovered by ${\cal A}_{D[R]}$ nor ${\cal A}_{D[\bar{R}]}$.

\begin{figure}[htbp]
  \centering
  \includegraphics[width=0.5\textwidth]{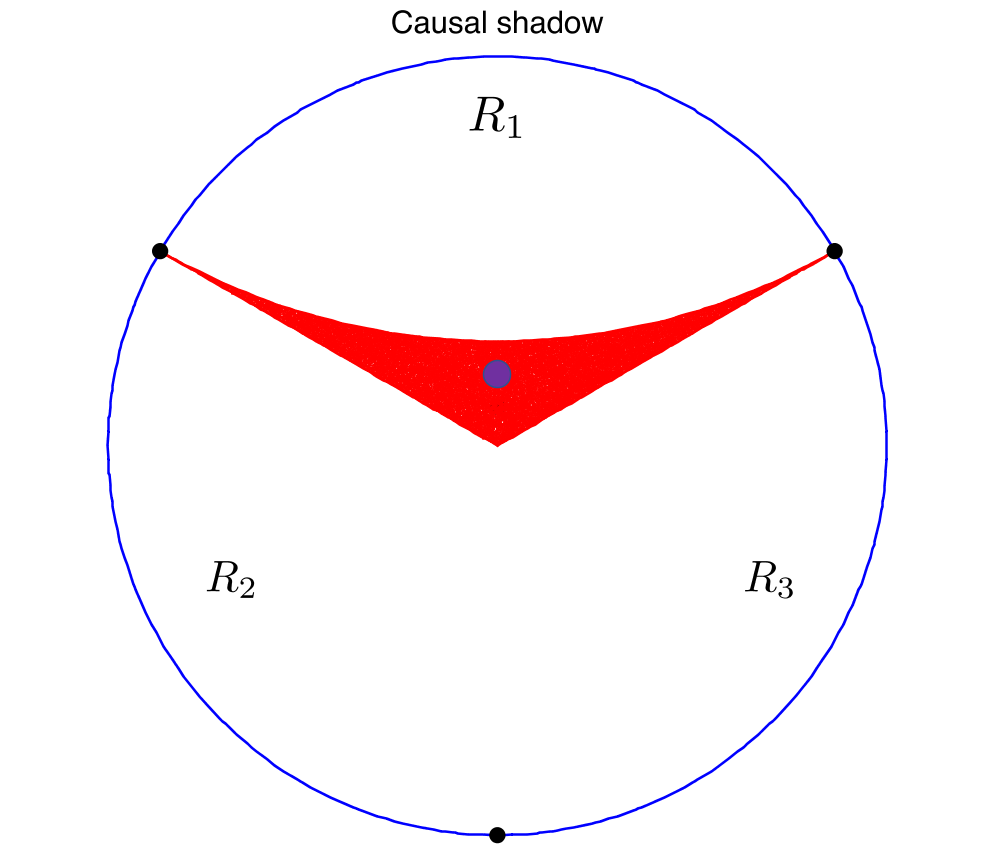}
  \caption{\label{fig4} A schematic plot for causal shadow. This figure is only illustrative and the causal wedges may not be precise.}
\end{figure}

The existence of the causal shadow is generic in the geometries we are considering; for instance, this occurs in dS/dS as illustrated in Fig.~\ref{fig2}. This indicates that there are some bulk points inside the code subspace that are protected both from erasing ${\cal A}_{D[R]}$ or ${\cal A}_{D[\bar{R}]}$.
This corresonds to them not being reconstructable using these algebras.  Moreover, if we take the three-partite example, see Fig.~\ref{fig4}, the region of the causal shadow $\text{CS}[R_1;\bar{R}_1]$ is shown in red. The operator of the purple point in the bulk could not be recovered by the HKLL formula by any of $R_1$, $R_2$, $R_3$ or $\bar{R}_1$, but it could be recovered by $\bar{R}_2$ and $\bar{R}_3$. 

On the other hand, we may use EW reconstruction, or a combination of EW and CW reconstruction, in our problem.  We could perhaps optimally work with the reconstruction $\max (\text{EW} [R], \text{CW}[R])$, namely, choosing the larger of the two bulk regions between CW and EW associated to a given boundary region $R$. Thus, in the dS/dS slicing, for $R$ smaller than half of the boundary, we may choose $\text{CW}[R]$ for $R$ and $\text{EW}[\bar{R}]$ for $\bar{R}$.  As we stressed in \S\ref{sec:overlap}, the intersection between $\text{CW}[R]$ and $\text{EW}[\bar{R}]$ is not empty. This means that quantum information in the causal shadow is reconstructable overall. This overlap is an interesting violation of geometric complementarity as defined in \cite{Pastawski:2016qrs}.

Given this violation  of geometric complementarity, it would be interesting to investigate the quantum no-cloning theorem in cutoff holography.\footnote{We thank K. Kato for asking about this, and J. Preskill for useful discussions.} The fact that one can recover the same state from both $R$ and $\bar{R}$ suggests that the encoding maps based on $\text{CW}[R]$ or $\text{EW}[\bar{R}]$ can be used to copy states. However, we are using two different maps which, due to (\ref{eq:novel2}), do not commute, and so the no-cloning theorem need not be violated. We hope to return to these issues in future work.

\subsection{Local correctability}

Let us now discuss local correctability, a stronger criterion for error correction codes than that they correct errors. We begin by reviewing the
condition studied in the recent work 
\cite{Pastawski:2016qrs}; it starts from the following structure which arises in a lattice system with a factorized Hilbert space defined on different spatial regions.  Consider a holographic error correction code and a connected region $R$ given by three disjoint intervals $R_1, H, R_2$. Now we consider erasing the middle region $H$. If we could erase it and could still recover the information in the bulk from $R'={R_1} \cup {R_2}$, then there is a recovery map $\mathcal{R}$ based on CW or EW reconstruction that could correct the error. Now we could demand a stronger condition: that the error in $H$ could be corrected by a recovery map that only acts on a region that is only a little bit larger than $H$. This is called local correctability and it is stronger than the usual error correcting criterion. 

Local correctability could be guaranteed by the quantum Markov condition \cite{Flammia:2016xvs}. As we discussed above in \S\ref{sec:bSSA}, for three disjoint regions $A$, $B$ and $C$, the quantum Markov condition says that 
\begin{align}\label{QMarkov}
0 = I(A;C|B) = S(AB) + S(BC) - S(ABC) - S(B)\,.
\end{align}
If this condition is satisfied, meaning that strong subadditivity is saturated, then there exists a recovery map $\mathcal{R}^{B,BC}$ that takes $\rho_{AB}$ towards $\rho_{ABC}$ known as the Petz map. In the holographic setting, we need to set $A=\bar{R}$, $B=R'$ and $C=H$, so that the Petz map if it exists would achieve the recovery while only acting on $R'$, i.e. acting locally near the excised region $H$. It is now straightforward to show that if the quantum Markov condition (\ref{QMarkov}) is satisfied, then
\begin{align}\label{simplerMarkov}
S(H) + S(R) = S\left( {R'} \right)\,.
\end{align}
One can check this in the large $c$ approximation to the entropy provided by Ryu-Takayanagi, something that we have reproduced on the deformed QFT side in the case studies we analyzed in this work.

The work in \cite{Pastawski:2016qrs} finds that generically local correctability is violated for flat space and a hemisphere, the latter being essentially our dS/dS case at $w_c=\pi\ell/2$. Here we will describe the status of this criterion in our two main case studies of cutoff dS/dS and AdS/Poincar\'e. In dS/dS geometry with boundary $w_c$ (\ref{mets}), we denote $\left| {{R_1}} \right| = \left| {{R_2}} \right| = \frac{r}{2}\Delta \phi$ and $|H|=(1-r) \Delta \phi$. Demanding the quantum Markov condition (\ref{simplerMarkov})
for $w_c=\pi/2$ would require
\be\label{dSlocfail}
(1-r)\Delta\phi + [r\Delta\phi+(1-r)\Delta\phi]=r\Delta\phi
\ee
since the entropy is given by the RT surface length, which is just the great circle along the boundary in this case.  This only allows $r=1$, i.e. $H=0$, implying the absence of local correctability \cite{Pastawski:2016qrs}.
One can analyze this similarly for $w_c<\pi\ell/2$.
Another example of this occurs in our other case study of cutoff AdS/Poincar\'e using our expression (\ref{eq:S2d}) for the entropy.  
This is likely generic; indeed the absence of local correctibility in the bulk flat space case \cite{Pastawski:2016qrs}\ applies to sufficiently strongly cut off (A)dS models, once the boundary goes beneath the (A)dS radius scale.
We expect that the violation of local correctability might be a generic feature  providing another probe of the nonlocality of the corresponding dual $T\bar T+\dots$ deformed quantum field theories.

\subsection{Quantum simulation}
The proposal of cutoff $\text{(A)dS}/T\bar{T}$ duality is not only useful for probing integrable trajectories deforming quantum field theory, but also helpful for understanding novel holographic dualities.   This is particularly relevant for de Sitter space and cosmology, where we would like to characterize connections between quantum gravity as it arises in nature, and quantum information. The study might benefit from the rapid development of quantum information technology (see, for instance \cite{preskill2018quantum,Preskill:2018fag}).\footnote{Related research includes time dependent brane dynamics \cite{Cooper:2018cmb,Antonini:2019qkt} or toy models for bubble collision and cosmic phase transition \cite{cosmology}.} 
We hope that this work might inspire future research for designing corresponding novel toy models of quantum error correction codes. Based on the results in this paper, the corresponding toy models should satisfy the following properties:
\begin{itemize}
\item  \emph{Finite} entanglement entropy given by the RT formula for entanglement entropy at large $c$ and large seed theory 'tHooft coupling. 
In this work, we obtained the derivative $L S'(L)$ explicitly on both sides of the duality.  It would be interesting to nail down the integration constant in the $2d$ description; the RT prescription yields a finite entropy.
\item Violation of the boosted SSA, indicating that new operators become accessible to a boosted observer (as in string theory) as described in \S\ref{sec:bSSA}.    
\item Violation of geometric complementarity if we consider both CW and EW reconstruction, i.e. the overlap described in \S\ref{sec:overlap}.
\item Violation of local correctability.
\end{itemize}
 The above list summarizes the level of non-locality that we find in the cutoff $\text{(A)dS}/T\bar{T}$ holography. We leave the development of constructing the corresponding quantum information theoretic codes to future research. 
 
\vskip 5mm  

\subsection{Final comments}

Finally, we note some potential connections to other recent developments in quantum gravity.
Recently, important progress has been made in another direction by deriving the Page curve for black hole evaporation using entanglement wedge reconstruction (for instance, see \cite{Penington:2019npb,Almheiri:2019psf,Almheiri:2019hni,Akers:2019nfi,Almheiri:2019yqk,Penington:2019kki,Almheiri:2019qdq}). 
It would be very interesting to understand if these results generalize to the deformed theories which characterize finite bulk patches of spacetime.  
As mentioned in \cite{Penington:2019npb}, combining these developments may help generalize the results to more realistic spacetimes (see also \cite{Penington:2019kki}).   Isolating a radially cutoff portion of the bulk seed AdS theory, in addition to being essential for the further deformation to a de Sitter bulk, may also facilitate decomposition of the system into different sectors (including the islands related to Hawking radiation).  

In general, there is a wide variety of important dynamical effects in realistic quantum gravity, some of them dominating over Euclidean quantum gravity contributions.  It would be interesting to understand the generalization of our results beyond large $c$, something that likely involves the decay of de Sitter to a more general FRW evolution \cite{Dong:2011uf}\ where the precision of the holographic description grows asymptotically at late times.

\noindent{\bf Acknowledgements}

\smallskip

\noindent We are grateful to J. Aguilera-Damia, H. Casini, J. Cardy, X. Dong, V. Gorbenko, M. Guica, K. Kato, A. Levine, D. Marolf, E. Mazenc, A. Milsted, Y. Nomura, M. Rangamani, P. Rath, B. \c{S}ahino\u{g}lu, N. Salzetta, E. Shaghoulian, W. Song, R. Soni, J. Sorce, H. Verlinde, G. Vidal and A. Wall for useful discussions. JL especially thanks C. Cheung, J. Preskill and D. Simmons-Duffin for their numerous discussions and support, and Stanford University for hosting during the course of the project.
We thank the organizers and participants of the Simons Center Workshop ``$T\bar T$ and other solvable deformations of quantum field theory'', the Yukawa Institute It from Qubit school and workshop, the Amsterdam summer workshop on string theory, and the Aspen Center for Physics where various parts of this work were completed.  
JL is supported in part by the Institute for Quantum Information and Matter (IQIM), an NSF Physics Frontiers Center (NSF Grant PHY-1125565) with support from the Gordon and Betty Moore Foundation (GBMF-2644), by the Walter Burke Institute for Theoretical Physics, and by Sandia Quantum Optimization \& Learning \& Simulation, DOE Award \#DE-NA0003525.
The research of E.S. was supported in part by the Simons Foundation Origins of the Universe Initiative (modern inflationary cosmology collaboration), by a Simons Investigator award, and by the National Science Foundation  under grant number PHY-1720397.  ES thanks the Aspen Center for Physics (NSF PHY-1607611) for hospitality during part of this work. GT is supported by CONICET (PIP grant 11220150100299), UNCuyo, and CNEA.  GT would like to acknowledge hospitality and support from the Aspen Center for Physics (NSF grant PHY-1607611, and Simons Foundation grant), and Stanford University, where part of this work was performed.

\appendix

\section{HKLL formula in dS/dS slicing}\label{sec:HKLLappendix}

As an example of the generalized HKLL\cite{Hamilton:2005ju,Hamilton:2006az} formula, we describe a reconstruction in dS/dS where the Dirichlet wall is at $\frac{\pi \ell}{2}$,
\begin{align}
d s_{d=3}^{2}=d w^{2}+\sin ^{2}\left(\frac{w}{\ell}\right)\left(-d \tau^{2}+\ell^{2} \cosh ^{2} \frac{\tau}{\ell} d \phi^{2}\right)~.
\end{align}
We consider a scalar field $\Phi$ moving in the $\text{dS}_3$ spacetime with a $\text{dS}_2$ slicing. The scalar field $\Phi$ follows the Klein-Gordon equation
\begin{align}
\left( {\frac{1}{{\sqrt { - g} }}{\partial _\mu }\left( {\sqrt { - g} {g^{\mu \nu }}{\partial _\nu }} \right) - {m^2}} \right)\Phi  = 0~,
\end{align}
where $m$ is the mass of the scalar field. We solve this equation by separation of variables. We assume
\begin{align}
\Phi  \propto \chi (w)\tilde \varphi (\tau ,\phi )~,
\end{align}
where we assume $\tilde{\varphi}$ could solve the Klein-Gordon equation in $d=2$, with the metric,
\begin{align}
d s_{d=2}^{2}=-d \tau^{2}+\ell^{2} \cosh ^{2} \frac{\tau}{\ell} d \phi^{2}~.
\end{align}
Thus we reduce the original equation to the following two equations
\begin{align}
&- \partial _\tau ^2\tilde \varphi  + \frac{1}{{{\ell ^2}}}{{\mathop{\rm sech}\nolimits} ^2}\left( {\frac{\tau }{\ell }} \right)\partial _\phi ^2\tilde \varphi  - \frac{1}{\ell }\text{tanh} \left( {\frac{\tau }{\ell }} \right){\partial _\tau }\tilde \varphi  - m_{d = 2}^2\tilde \varphi  = 0~,\nonumber\\
&\partial _w^2\chi  + \frac{2}{\ell }\cot \left( {\frac{w}{\ell }} \right){\partial _w}\chi  + {\csc ^2}\left( {\frac{w}{\ell }} \right)m_{d = 2}^2\chi  - {m^2}\chi  = 0~,
\end{align}
where $m_{d=2}$ is the mass from the Klein-Gordon equation in $d=2$. Here we choose the following solution
\begin{align}
{\Phi _{n,\mu}}(w,\tau ,\phi ) = C_{\mu}{\chi_\mu }(w){y_{n,\mu }}(\tau ){S_n}(\phi )~,
\end{align}
where $\lambda,n$ are labelling different modes. $\mu$ is continuous and $n$ is discrete. The functions and constants are defined by
\begin{align}
&{C_\mu } = \frac{1}{{\sqrt {X(\mu ,\nu )} }}~,\nonumber\\
&{\chi _\mu } = \frac{1}{{\sqrt \ell  }}\frac{{P_\nu ^{i\mu }\left( {\cos \left( {\frac{w}{\ell }} \right)} \right) + \gamma _\nu ^\mu Q_\nu ^{i\mu }\left( {\cos \left( {\frac{w}{\ell }} \right)} \right)}}{{\sqrt {\sin \left( {\frac{w}{\ell }} \right)} }}~,\nonumber\\
&{y_{n,\mu }}(\tau ) = \frac{{{2^n}}}{{\sqrt \mu  }}{\cosh ^n}\left( {\frac{\tau }{\ell }} \right){e^{\left( {n + \frac{1}{2} - i\mu } \right)\frac{\tau }{\ell }}}_2{F_1}\left( {n + \frac{1}{2},n + \frac{1}{2} - i\mu ;1 - i\mu ; - {e^{\frac{{2\tau }}{\ell }}}} \right)~,\nonumber\\
&S_{n}(\phi)=\frac{1}{\sqrt{2 \pi}} e^{i n \phi} ~~~~~~n\in \mathbb{Z}~,
\end{align}
where
\begin{align}
&\mu  = \sqrt {m_{d = 2}^2{\ell ^2} - \frac{1}{4}} ~,\nonumber\\
&\nu=\sqrt{1-m^{2} \ell^{2}}-\frac{1}{2}~,\nonumber\\
&\gamma _\nu ^\mu  =  - \frac{{4\csc \left( {\frac{1}{2}\pi (i\mu  + \nu )} \right)\Gamma \left( {\frac{1}{2}( - i\mu  + \nu  + 2)} \right)}}{{\Gamma \left( {\frac{1}{2}( - i\mu  - \nu )} \right)\Gamma \left( {\frac{1}{2}( - i\mu  + \nu  + 1)} \right)\Gamma \left( {\frac{1}{2}(i\mu  + \nu  + 1)} \right)}}~,\nonumber\\
&X(\mu ,\nu ) \equiv {E_{PP}}(\mu ,\nu ) + 2{\mathop{\rm Re}\nolimits} \gamma _\nu ^{\mu *}{E_{PQ}}(\mu ,\nu ) + {\left| {\gamma _\nu ^\mu } \right|^2}{E_{QQ}}(\mu ,\nu )~,\nonumber\\
&{E_{PP}}(\mu ,\nu )\equiv \frac{\pi }{{\Gamma (1 - i\mu )\Gamma (1 + i\mu )}} + \frac{{{{\sin }^2}\left( {\pi \nu } \right)\Gamma (i\mu )\Gamma ( - i\mu )}}{\pi }\nonumber\\
&+ \frac{{\pi \Gamma (i\mu )\Gamma ( - i\mu )}}{{\Gamma (1 + \nu  - i\mu )\Gamma ( - \nu  - i\mu )\Gamma (1 + \nu  + i\mu )\Gamma ( - \nu  + i\mu )}}~,\nonumber\\
&{E_{QQ}}(\mu ,\nu ) \equiv \frac{{{\pi ^3}}}{{4{{\sinh }^2}(\pi \mu )}}\left[ {\frac{{{{\cosh }^2}(\pi \mu ) + 1}}{{\Gamma (1 - i\mu )\Gamma (1 + i\mu )}} + \Gamma (i\mu )\Gamma ( - i\mu )\left( {{A_\mu }{A_{ - \mu }} + {B_\mu }{B_{ - \mu }}} \right)} \right]~,\nonumber\\
&{E_{PQ}}(\mu ,\nu ) \equiv \frac{{i\pi }}{{2\sinh (\pi \mu )}}\left\{ {\begin{array}{*{20}{l}}
{\frac{{2\sin (\pi \nu )\Gamma (i\mu )\Gamma ( - i\mu )}}{{\Gamma (1 + \nu  + i\mu )\Gamma ( - \nu  - i\mu )}}}\\
{ + \cosh (\pi \mu )\left[ {\begin{array}{*{20}{l}}
{\frac{\pi }{{\Gamma (1 - i\mu )\Gamma (1 + i\mu )}} + \frac{{{{\sin }^2}(\pi \nu )\Gamma (i\mu )\Gamma ( - i\mu )}}{\pi }}\\
{ + \frac{{\pi \Gamma (i\mu )\Gamma ( - i\mu )}}{{\Gamma (1 + \nu  - i\mu )\Gamma ( - \nu  - i\mu )\Gamma (1 + \nu  + i\mu )\Gamma ( - \nu  + i\mu )}}}
\end{array}} \right]}
\end{array}} \right\}~,
\end{align}
and the function $P$ and $Q$ are generalized Legendre functions, which are related to the hypergeometric function by
\begin{align}
&P_\nu ^\mu (x) = \frac{1}{{\Gamma (1 - \mu )}}{\left[ {\frac{{1 + x}}{{1 - x}}} \right]^{\mu /2}}{_2F_1}( - \nu ,\nu  + 1;1 - \mu ;\frac{{1 - x}}{2})~,\nonumber\\
&Q_\nu ^\mu (x) = \frac{{{\pi ^{1/2}}}}{{{2^{\nu  + 1}}}}\frac{{\Gamma (\mu  + \nu  + 1)}}{{\Gamma (\nu  + 3/2)}}\frac{1}{{{x^{\mu  + \nu  + 1}}}}{(1 - {x^2})^{\mu /2}}{_2F_1}(\frac{{1 + \mu  + \nu }}{2},\frac{{2 + \mu  + \nu }}{2},\nu  + \frac{3}{2};\frac{1}{{{x^2}}})~.
\end{align}
The choice of the above wavefunction satisfies
\begin{itemize}
\item Consistent normalization. The normalization is fixed by the Klein-Gordon inner product
\begin{align}
\left( {{\Phi _{n,\mu}},{\Phi _{\bar{n},\bar{\mu}}}} \right) =  - i\int_\Sigma  {d{\Sigma ^a }\left( {{\phi _{n,\mu}}{{\mathord{\buildrel{\lower3pt\hbox{$\scriptscriptstyle\leftrightarrow$}} 
\over \partial } }_a }\phi _{\bar{n},\bar{\mu}}^*} \right)}  =\delta _{n,\bar{n}}\delta(\mu-\bar{\mu})~,
\end{align}
where $\Sigma$ is an arbitrary time slice and the indices $a$ are labelling spatial coordinates. The normalization is fixed by some integral formulas of $P$ and $Q$ used in \cite{Nguyen:2017ggc,bielski2013orthogonality}. Here we assume that $\mu\in \mathbb{R}$.
\item Dirichlet wall. The condition $\Phi(w=\pi\ell/2)=0$ is satisfied, in order to fix the ratio between the prefactors of $P$ and $Q$. 
\end{itemize}
According to the above quantization, we could write down a formal form of the HKLL formula, using the language of \cite{Bousso:2012mh}. Denoting boundary coordinates by $x$ and bulk coordinates by $X$, the scalar field quantization for the bulk field $\Phi$ and the boundary field $\phi$ are given by
\begin{align}
&\Phi (X) = \int d k{a_k}{F_k}(X)~,\nonumber\\
&\phi (x) = \int d k{a_k}{f_k}(x)~.
\end{align}
Here for simplicity, we only write down the annihilation operator part, and the full expression should contain its conjugates. In our case, we also have $k=(n,\mu)$ and $\int dk\equiv \sum_n \int d\mu$. Note that
\begin{align}
\int d x\varphi (x)f_p^*(x) = \int d x\int d k{a_k}{f_k}(x)f_p^*(x) = \int d k{V_{kp}}{a_k}~.
\end{align}
If the matrix 
\begin{align}
V_{k p} \equiv \int d x f_{k}(x) f_{p}^{*}(x)~,
\end{align}
could be inverted, namely, we could find $V^{-1}$ such that 
\begin{align}
\int {dp} {V_{kp}}{(V)^{ - 1}}_{pq} = {\delta _{kq}}~,
\end{align}
then we have
\begin{align}
{a_q} = \int {dp} {(V)^{ - 1}}_{pq}\int d x\varphi (x)f_p^*(x)~.
\end{align}
Then we find the existence of the HKLL formula
\begin{align}
&\Phi (X) = \int d k{a_k}{F_k}(X) = \int d k\int {dp} {(V)^{ - 1}}_{pk}\int d xf_p^*(x)\varphi (x){F_k}(X)\nonumber\\
&\equiv \int {dbK(X|x)\varphi (x)}~,
\end{align}
where
\begin{align}
K(X|x) = \int d kdpf_p^*(x){(V)^{ - 1}}_{pk}{F_k}(X)~.
\end{align}
Specifying our case, we have $X=(w,x)$ and $x=(\tau,\phi)$. The function $F$, $f$ and $V$ are listed as
\begin{align}
&{F_{n,\mu }}(X) = {C_\mu }{\chi _\mu }(w){y_{n,\mu }}(\tau ){S_n}(\phi )~,\nonumber\\
&{f_{n,\mu }}(x) = {C_\mu }{\partial _w}{\chi _\mu }\left( {w = \frac{{\pi \ell }}{2}} \right){y_{n,\mu }}(\tau ){S_n}(\phi ) \equiv {C_\mu }\xi (\mu ,\nu ){y_{n,\mu }}(\tau ){S_n}(\phi )~,\nonumber\\
&{\partial _w}{\chi _\mu }\left( {w = \frac{{\pi \ell }}{2}} \right) = \frac{1}{{{\ell ^{3/2}}}}\sqrt \pi  {2^{1 + i\mu }} \times \nonumber\\
&\frac{{\left( {\Gamma \left( {\frac{1}{2}( - i\mu  + \nu  + 1)} \right)\Gamma \left( {\frac{1}{2}(i\mu  + \nu  + 1)} \right) + 2\cot \left( {\frac{1}{2}\pi (\nu  + i\mu )} \right)\Gamma \left( {\frac{1}{2}( - i\mu  + \nu  + 2)} \right)\Gamma \left( {\frac{1}{2}(i\mu  + \nu  + 2)} \right)} \right)}}{{\Gamma \left( {\frac{1}{2}( - i\mu  - \nu )} \right)\Gamma {{\left( {\frac{1}{2}( - i\mu  + \nu  + 1)} \right)}^2}\Gamma \left( {\frac{1}{2}(i\mu  + \nu  + 1)} \right)}}\nonumber\\
&\equiv \xi (\mu ,\nu )~,\nonumber\\
&{V_{n,\mu ;\bar n,\bar \mu }} = {C_\mu }C_{\bar \mu }^*\xi (\mu ,\nu ){\xi ^*}(\bar \mu ,\nu ){\delta _{n\bar n}}\int d \tau {y_{n,\mu }}(\tau )y_{n,\bar \mu }^*(\tau )~,
\end{align}
and the HKLL kernel is given by
\begin{align}
K(X | x)=\sum_{n, \overline{n}} \int d \mu d \overline{\mu} f^*_{n, \mu}(x)\left(V^{-1}\right)_{n, \mu, \overline{n}, \overline{\mu}} F_{\overline{n}, \overline{\mu}}(X)~.
\end{align}
In this dS/dS slicing example we are studying, we find it is not easy to obtain a full analytic form of the HKLL kernel. Due to the study of \cite{Morrison:2014jha}, the HKLL kernel should be understood as a formal distribution in the sense of algebraic quantum field theory, considering that this integral might be divergent.

This is an example showing that the HKLL formula in the geometries we are studying in the main text, is analytically trackable. We leave more detailed studies of the HKLL formula in Fourier space, or other patches, or cutoff AdS to future works.  
 
\section{Characteristics for Stress Tensor PDEs and additional examples}\label{sec:Charappendix}

As discussed in the main text, it is of interest for various tests and applications of the dualities to understand the dressed stress energy as defined by the trace flow and conservation equations (\ref{Teqs}).  After solving for one of the three components of $T_{ab}$ using the trace flow equation, the conservation equations take the form of a system of two quasilinear PDEs for two functions~\cite{characteristics}, which we will denote by the 2-vector $\vec U$.

In any system of coordinates $x_1, x_2$, the PDEs take the form
\be\label{Genpde}
\partial_{x_1} \vec U + B(\vec U, x_1, x_2) \partial_{x_2}\vec U=\vec C(\vec U, x_1, x_2)
\ee
with the $2\times 2$ matrix $B$ and 2-vector $\vec C$ depending on the coordinates and $\vec U$ (but not depending on derivatives of $\vec U$).  

If the eigenvalues $\lambda_i$ of $B\equiv P\Lambda P^{-1}, \Lambda=\text{diag}(\lambda_1, \lambda_2)$ are real, the PDEs define real characteristic curves.  This corresponds to either hyperbolic or parabolic PDEs; if the characteristics are complex the system is elliptic.   These are useful for constructing solutions by tracing out the behavior of $\vec U$ on each characteristic.   This yields a well defined problem if we specify initial data on a curve which crosses each characteristic once. 

We define two families of characteristics as~\cite{characteristics}
\be\label{charsgen}
\frac{dx_1}{ds_i}=1, ~~~ \frac{dx_2}{ds_i}=\frac{dx_2}{dx_1}|_i=\lambda_i
\ee
with the functions $u_1, u_2$ evolving according to the equation
\be\label{Uchars}
\sum_{j=1}^n P^{-1}_{ij} \frac{d u^{(j)}}{ds_i}=\sum_{j=1}^n P^{-1}_{ij} C_j\,.
\ee

In the main text, in section \S\ref{sec:dScharspacman}\ we analyze this for dS/dS, finding a useful set of real characteristics for the case $w_c=\pi\ell/2$. 
For the remainder of this appendix, we will similarly analyze the case of a flat 2d spacetime.  This is a simple example which will also provide a consistency check of the $\lambda\to0$ limit.

In Euclidean flat coordinates $d\tau^2+dx^2$ (which could apply to either the cylinder or Poincar\'e, depending on whether we compactify $x$), we have the conservation equations
\bea\label{consflat}
\partial_\tau T^\tau_\tau +\partial_x T^x_\tau &=& 0 \nonumber\\
\partial_\tau T^\tau_x + \frac{\partial T^x_x}{\partial T^x_\tau}\partial_x T^x_\tau + \frac{\partial T^x_x}{\partial T^\tau_\tau}\partial_x T^\tau_\tau &=& 0
\eea
where we use the trace flow equation to eliminate $T^x_x$.  Defining the shorthand
\be\label{shorthand}
T^\tau_\tau = u_1, ~~~ T^x_\tau = u_2, ~~~ \tilde\lambda = \pi\lambda, ~~~ c_2=1-\eta
\ee
the trace flow equation is
\be\label{TxxTFE}
T^x_x(u_1, u_2)=-\frac{u_1+\tilde\lambda u_2^2-c_2/\tilde\lambda}{1-\tilde\lambda u_1}
\ee
at the level of pure gravity.  
Defining $\vec U=(u_1, u_2)$, our system of PDEs is
\be\label{Consmatrix2}
\partial_\tau \vec U + B \partial_x\vec U=0
\ee
with
\be\label{Bmatrix} 
   B=
  \left[ {\begin{array}{cc}
   0 & 1 \\
   \partial_{u_1}T^x_x & \partial_{u_2}T^x_x \\
  \end{array} } \right]\,.
\ee
It is straightforward to calculate the eigenvalues and eigenvectors of this matrix $B$, diagonalizing it as $B=P\Lambda P^{-1}$
with $\Lambda=\text{diag}(\lambda_1, \lambda_2)$. 
In the present case of flat geometry, we have zero on the RHS of (\ref{Consmatrix2})\ and the equations
for the two sets of characteristics labeled by $i=1,2$ is
\be\label{chareqs}
\frac{d\tau}{ds_i}=1, ~~~ \frac{dx}{d\tau}|_i=\lambda_i, ~~~\sum_{j=1}^2 P^{-1}_{ij} \frac{d}{ds_i} u^{(j)}=0\,.
\ee 
We find that B has eigenvalues
\be\label{evaluesflat}
\lambda_\pm = \frac{1}{(1-\tilde\lambda u_1)}\left(\tilde\lambda u_2 \pm \sqrt{c_2-1} \right)=\lambda_\pm = \frac{1}{(1-\tilde\lambda u_1)}\left(\tilde\lambda u_2 \pm \sqrt{-\eta} \right)\,.
\ee
These enter into (\ref{chareqs}).  

We note that in the pure CFT case, $c_2=1-\eta=0=\tilde\lambda$, these eigenvalues are always complex, corresponding to elliptic PDEs. It is a (minor) check to see that the solutions are indeed smooth in that limiting case. This persists to $\tilde\lambda\ne 0$  
for any $c_2<1$.  In holographic theories, this corresponds to the bulk-AdS$_3$ case.  

Next, we notice that in contrast, the $\Lambda_2$ deformation corresponding to bulk 3d Minkowski space ($c_2=1$) or bulk $\text{dS}_3$ ($c_2>1$) has 
real eigenvalues for all values of $u_1, u_2$.  In these cases, each of the two sets of characteristics are generically real.  

The nonlinearities of the PDEs -- which were introduced by the deformation via the trace flow equation -- may be expected to generically lead the characteristics to cross, indicating discontinuities.  
We leave a more detailed investigation of this flat 2d case and its implications to future work, focusing on the dS/dS $w_c=\pi\ell/2$ example in the main text.
For now, we just note the precise correspondence we obtained from the simple formula (\ref{evaluesflat}) between the bulk geometry in the holographic case and the ellipticity (or not) of the boundary theory stress energy equations (\ref{Teqs}).

\bibliography{dS}{}
\bibliographystyle{utphys}

\end{document}